\documentclass[11pt]{article}

\usepackage{verbatim}
\usepackage{amsmath,amscd}
\usepackage{amsfonts}
\usepackage{amssymb}
\usepackage{bbm}
\usepackage{latexsym}
\usepackage{graphics}
\usepackage{lscape} 
\usepackage{epsfig}
\usepackage{color}
\usepackage{longtable}
\usepackage{pdflscape}

\usepackage{amscd}
\usepackage{cite}

\newcommand{\Z}[1]{\ensuremath{\mathbbm{Z}_{#1}}} 
\newcommand{\Id}{\ensuremath{1\!\!1}}
\newcommand{\TabSp}{$\phantom{I^{I^I}}$\!\!\!\!}

\DeclareMathSymbol{\mg}{\mathrel}{symbols}{"1D}

\newcommand{\gb}{\beta}

\newcommand{\gth}{\theta}

\newcommand{\go}{\omega}

\newcommand{\gp}{\pi}

\newcommand{\gch}{\chi}
\newcommand{\gL}{\Lambda}

\newcommand{\cF}{{\cal F}}

\newcommand{\cR}{{\cal R}}

\newcommand{\tN}{{\widetilde N}}

\newcommand{\tr}{\text{tr}}

\newcommand{\ra}{\rightarrow}

\newcommand{\dsp}{\displaystyle}

\newcommand{\beq}{\begin{equation}}
\newcommand{\eeq}{\end{equation}}
\newcommand{\barr}{\begin{array}}
\newcommand{\earr}{\end{array}}
\newcommand{\equ}[1]{\begin{gather} #1 \end{gather}}
\newcommand{\equa}[1]{\begin{align} #1 \end{align}}

\newcommand{\arry}[2]{\begin{array}{#1} #2 \end{array}}

\newcommand{\sfrac}[2]{\mbox{$\frac{#1}{#2}$}}
\newcounter{oldcounter}

\newcommand{\tga}{{\tilde \alpha}}

\newcommand{\Intr}{\mathbb{Z}}
\newcommand{\Cplx}{\mathbb{C}}
\newcommand{\Real}{\mathbb{R}}

\newcommand{\ba}[2]{\[\begin{array}{#2}\label{#1}}
\newcommand{\ea}{\end{array}\]}
\newcommand{\be}{\begin{equation}}
\newcommand{\ee}{\end{equation}}
\newcommand{\bea}{\begin{eqnarray}}
\newcommand{\eea}{\end{eqnarray}}

\newcommand{\E}[1]{\mathrm{E_{#1}}}
\newcommand{\U}[1]{\mathrm{U(#1)}}
\newcommand{\SU}[1]{\mathrm{SU(#1)}}
\newcommand{\SO}[1]{\mathrm{SO(#1)}}

\newcommand{\rep}[1]{\mathbf{#1}}
\newcommand{\crep}[1]{\mathbf{\overline{#1}}}

\newcommand{\sm}{{\,\mbox{-}}}

\begin{document}

\thispagestyle{empty}

\begin{flushright}
DESY-12-242 \\
LMU-ASC 86/12 
\\
\end{flushright}
\vskip 2 cm
\begin{center}
{\Large {\bf 
 Schoen manifold with line bundles as resolved magnetized orbifolds
} 
}
\\[0pt]

\bigskip
\bigskip {\large
{\bf Stefan Groot Nibbelink$^{a,}$}\footnote{
E-mail: Groot.Nibbelink@physik.uni-muenchen.de},
{\bf Patrick K.S. Vaudrevange$^{b,}$}\footnote{
E-mail: patrick.vaudrevange@desy.de}
\bigskip }\\[0pt]
\vspace{0.23cm}
${^a}$ {\it 
Arnold Sommerfeld Center for Theoretical Physics,\\
~~Ludwig-Maximilians-Universit\"at M\"unchen, 80333 M\"unchen, Germany
 }\\[1ex]  
 ${}^b$ {\it 
Deutsches Elektronen--Synchrotron DESY, Notkestra\ss e 85, 22607 Hamburg, Germany
}

\bigskip
\end{center}

\subsection*{\centering Abstract}

We give an alternative description of the Schoen manifold as the blow--up of a 
$\Z{2}\times\Z{2}$ orbifold in which one $\Z{2}$ factor acts as a 
roto--translation. Since for this orbifold the fixed tori are only identified 
in pairs but not orbifolded, four--dimensional chirality can never be obtained 
in heterotic string compactifications using standard techniques alone. 
However, chirality is recovered when its tori become magnetized. To exemplify 
this, we construct an $\E{8}\times\E{8}'$ heterotic $\SU{5}$ GUT on the 
Schoen manifold with Abelian gauge fluxes, which becomes an MSSM with three 
generations after an appropriate Wilson line is associated to its freely acting 
involution. We reproduce this model as a standard heterotic orbifold CFT 
of the (partially) blown down Schoen manifold with a magnetic flux. 
Finally, in analogy to a proposal for non--perturbative heterotic 
models by Aldazabal et al.\ we suggest modifications to the heterotic orbifold 
spectrum formulae in the presence of magnetized tori.

\newpage 
\setcounter{page}{1}
\setcounter{footnote}{0}
%
%

\section{Introduction and summary} 
\label{sc:Intro}

There are two standard approaches in the literature to geometrical string 
compactification of the heterotic string. (Non--geometrical approaches involve 
e.g.\ free--fermionic models \cite{Faraggi:1991jr,Cleaver:1998sa} and Gepner 
constructions \cite{Dijkstra:2004cc,Dijkstra:2004ym}.) Either strings are 
considered on singular (toroidal) orbifolds or on smooth Calabi--Yau manifolds. 
The main advantage of orbifolds over smooth Calabi--Yau spaces is that they are 
so simple that the heterotic string can be quantized on them exactly 
\cite{Dixon:1985jw,Dixon:1986jc}. Therefore, one has access to the full 
spectrum of the theory; not just to its zero modes. In addition, one can 
scan in a very systematic way through the parameter space of heterotic orbifold 
compactifications in order to search for interesting models for string 
phenomenology (using e.g.\ \cite{Nilles:2011aj}). This has resulted, for 
example, in a mini--landscape of a few hundred MSSM models based on the 
heterotic $\E{8}\times\E{8}'$ 
orbifold $T^6/\Intr_\text{6--II}$ \cite{Lebedev:2006kn,Lebedev:2008un}.

An orbifold can be considered as a Calabi--Yau space at a singular point in its 
moduli space where symmetries get enhanced. To go away from the orbifold point 
in moduli space the orbifold singularities have to be resolved (or deformed). 
In this blow--up process certain (exceptional) cycles that were hidden inside 
the singularities acquire finite volumes. From the heterotic orbifold model perspective 
this corresponds to turning on Vacuum--Expectation--Values (VEVs) for twisted 
states, so--called blow--up modes, which are localized at the singularities of 
the orbifold. Unfortunately, an exact string quantization is out of reach at a 
generic point in moduli space and there is typically much less symmetry. 
For example, it turns out that in full blow--up any mini--landscape model has 
broken hypercharge \cite{Nibbelink:2009sp,Buchmuller:2012mu}. This might be 
interpreted in two ways: Either one does not go to the full blow--up in order to
keep hypercharge unbroken and our string vacuum is very close to the orbifold 
point, or our string vacuum is at a generic point of the moduli space and 
different constructions are needed for phenomenology. As discussed in 
\cite{Blaszczyk:2009in,Blaszczyk:2010db} freely acting involutions can be used 
as an example for the second interpretation and an MSSM orbifold model has been 
constructed on the heterotic $T^6/\Intr_2\times\Intr_2$ orbifold, which in 
principle can avoid hypercharge breaking in full blow--up.

Furthermore, there have been various constructions of MSSM models in the context 
of the heterotic string compactified on smooth Calabi--Yau manifolds. For 
example, a three generation MSSM has been constructed in \cite{Bouchard:2005ag} 
on the Schoen manifold \cite{Schoen:1988} using a stable $\SU{5}$ vector bundle 
\cite{Donagi:2000zf,Donagi:2000fw,Donagi:2004ub}. Similar constructions -- yet 
not fully supersymmetric \cite{Gomez:2005ii} -- can be 
found in e.g.\ \cite{Braun:2005bw}. Even though the Schoen manifold is just one particular 
Calabi--Yau space, it is a typical example of a complete intersection Calabi--Yau: 
It can be obtained as a set of hyper surfaces within a direct product of projective spaces.

Most heterotic models built on the Schoen manifold require complicated constructions of 
stable $\SU{N}$ bundles. Therefore, one may wonder whether it is also 
possible to design MSSM--like heterotic string models on the Schoen manifolds using line 
bundles. As has been realized by various groups \cite{Anderson:2011ns,Anderson:2012yf} 
the analysis of line bundles on smooth Calabi--Yau spaces, described as complete 
intersections in toric varieties, can be performed much easier than their 
non--Abelian counterparts. The main reason for this is that for line bundle 
gauge backgrounds the stability of the bundle reduces to solving simple 
Donaldson--Uhlenbeck--Yau (DUY) equations \cite{Donaldson:1985,Uhlenbeck:1986} 
in terms of the K\"ahler moduli \cite{Blumenhagen:2005pm,Blumenhagen:2005ga}. 
Moreover, the embedding of line bundles into the ten--dimensional heterotic gauge 
group ($\E{8}\times\E{8}'$ or $\SO{32}$, where we focus on the 
$\E{8}\times\E{8}'$ case, but most of our results equally apply to the 
$\SO{32}$ case.) can be characterized by vectors of integers \cite{Honecker:2006qz,Honecker:2007uw}. 
This makes it possible to perform computer--aided scans for potentially 
phenomenologically viable models.

The Schoen manifold does not only provide an interesting example of a 
Calabi--Yau constructed as a complete intersection. It can also be considered 
as a smooth limit of a certain orbifold \cite{Donagi:2008xy}. This orbifold has 
some special properties: It is a $T^6/\Intr_2 \times\Intr_\text{2,rototrans}$ 
orbifold, where $\Intr_\text{2,rototrans}$ acts as a roto--translation, i.e.\ 
as a simultaneously performed rotation and translation \cite{Hebecker:2004ce} 
(and \cite{Blumenhagen:2006ab} in the type II string context, where this kind of 
orbifolds are called shift orbifolds). 
This has far reaching consequences for the structure of the fixed points and 
tori and, in turn, modifies the breaking of higher dimensional supersymmetry 
to $\mathcal{N}=1$ in four dimensions for heterotic string compactifications. 
As we will see, this necessarily results in vector--like spectra for this kind 
of orbifold geometry.

This is not a peculiar feature of this special orbifold, many more orbifolds with 
this property are known. Recently, there has been a classification of all 
six--dimensional toroidal orbifold geometries that give rise to 
four--dimensional $\mathcal{N}=1$ supersymmetry \cite{Fischer:2012qj}. These 
geometries can be arranged in two sets: The ones with Abelian point group and 
the ones with non--Abelian point group. $23$ of the $138$ geometries with Abelian 
point group share the property that they necessarily lead to non--chiral 
spectra for heterotic string compactifications. (These are $\Z{2}\times\Z{2}$ variants, 
part of the classification of Ref.\ \cite{Donagi:2008xy}.)
For the non--Abelian cases these numbers are essentially unknown. However, one explicit 
example of a heterotic $S_3$ orbifold \cite{Konopka:2012gy} also turns out to produce only 
vector--like spectra. We therefore expect that also a sizable portion of the 
non--Abelian point group orbifolds will unavoidably be non--chiral in four dimensions.

Hence, it is an important question whether there exists an unavoidable no--go 
theorem against four--dimensional chirality for all these heterotic orbifolds. 
Fortunately, we will show that it is possible to circumvent this no--go by 
allowing for magnetized tori on the orbifold. Concretely, we put magnetic 
fluxes on the tori of the Schoen orbifold and show that four--dimensional 
chiral spectra can be realized. More than that, we will show that it is even 
possible to obtain MSSM--like models in this way.

There is one technical subtlety in the construction of such orbifolds with 
magnetized tori: As far as we know, contrary to conventional orbifolds, it is 
unknown how to quantize the heterotic string exactly on them. We by--pass this 
obstruction in two ways: 
 First, we consider the whole construction in blow--up, i.e.\ on the 
      smooth Schoen manifold.
 Second, we show that one can start with a six--dimensional spectrum 
      obtained from a standard heterotic $T^4/\Z{2}$ orbifold, which is a subspace of the 
      partially blown--down Schoen manifold, using conventional CFT techniques. 
      Then, one can use field theoretical methods, discussed e.g.\ in 
      \cite{Cremades:2004wa,Abe:2009vi,Abe:2009uz}, to determine the 
      consequences of the additional (magnetic) fluxes and to obtain a chiral 
      spectrum in four dimensions. 
Both approaches, i.e.\ the smooth approach and the hybrid approach of combining 
CFT and field theoretical methods, will reproduce exactly the same spectrum.

\subsection*{Paper overview}

In Section \ref{sc:Orbifolds} we review the basics of heterotic orbifold models. 
In addition we introduce the DW(0--2) orbifold which is of central interest in 
this work. Section \ref{sc:Schoen} provides an alternative description of the 
Schoen manifold as the resolution of this DW(0--2) orbifold. In Section 
\ref{sc:BundlesOnSchoen} we describe the $\E{8}\times\E{8}'$ heterotic 
string with line bundles on the divisors of the Schoen 
manifold including magnetic fluxes on the tori of the underlying orbifold. 
Moreover, we identify the relevant consistency conditions for such gauge 
backgrounds and compute the resulting chiral spectra in both, four and six, 
dimensions. Then, we provide an example that mainly serves to illustrate various 
aspects of the general theory developed in this paper. In Section 
\ref{sc:Models} we construct a specific example, which is potentially 
phenomenologically interesting as it has the particle spectrum of the MSSM in 
four dimensions. We analyze this example using two approaches: First, the 
smooth approach and, second, the hybrid approach of combining CFT and field 
theoretical methods. Finally, in Section \ref{sc:OrbifoldCFT} 
we speculate on how to extend the standard heterotic CFT description of 
orbifolds in the presence of magnetized tori.

\subsection*{Acknowledgements}

We would like to thank Vincent Bouchard and Ron Donagi for early discussions 
that initiated this project. We are also indebted to Kang--Sin Choi, James Gray, 
Tatsuo Kobayashi and Fabian R\"uhle for valuable discussions. 
SGN would like to thank the organizers of the Workshop``Topological Heterotic 
Strings and (0,2) Mirror Symmetry" in Vienna for hospitality. We also thank the 
organizers of the Bethe Forum and the 4th Bethe Center Workshop on ``Unification 
and String Theory" in Bonn/Bad Honnef for hospitality.
This research has been supported by the "LMUExcellent" Programme. P.V.\ is supported by SFB grant 676.

\section{Heterotic $\Z{2}\times\Z{2}$ orbifolds}
\label{sc:Orbifolds}

In this section we describe some basic geometrical properties of 
$\Intr_2\times\Intr_2$ orbifolds and explain how to determine whether such 
orbifolds can lead to heterotic ($\E{8}\times\E{8}'$) string models with chiral 
spectra in four dimensions. We follow the 
classification scheme for these orbifolds developed by Donagi--Wendland 
\cite{Donagi:2008xy}. (See their Table 1 for details and nomenclature). 
In particular, we describe their DW(0--2) orbifold which can be considered as a 
certain singular limit of the so--called Schoen manifold. However, for 
comparison purposes we first recall some basic facts of $\Intr_2\times\Intr_2$ 
orbifolds and give some details of the more often considered DW(0--1) orbifold.

\subsection{General features of $\Intr_2\times\Intr_2$ orbifolds}
\label{sc:GeneralFeatures}

We consider $\Z{2}\times\Z{2}$ orbifolds defined as 
\begin{equation}
\Real^6/S
\end{equation}
where the {\it space group} $S$ specifies an equivalence relation on $\Real^6$ 
as $g\,X \sim X$ for all $g \in S$ and $X \in \Real^6$. A general space group 
element $g = (\vartheta,\ell)$ consists of a six--dimensional rotation matrix 
$\vartheta$ and a translation $\ell$. It acts on $X \in \Real^6$ as 
$g\,X = \vartheta\, X + \ell$. The space group is generated by two types of 
elements: The purely translational elements $g_i = (\Id, e_i)$ are determined 
by six basis vectors $e_i$ ($i = 1,\ldots,6$) that span a six--dimensional 
lattice and hence define a six--torus. For simplicity, we identify 
$\Real^6 = \mathbbm{C}^3$ and take as basis vectors 
\equ{
e_1 = (1,0,0)~, \quad e_2 = (i,0,0)~, 
\quad 
e_3 = (0,1,0)~, \quad e_4 = (0,i,0)~, 
\quad 
e_5 = (0,0,1)~, \quad e_6 = (0,0,i)~. 
}
Consequently, we denote the torus coordinates by $z=(z_1,z_2,z_3) \in 
T^2_1 \times T^2_2 \times T^2_3$  in this complex basis. The remaining two 
generators of the space group, $g_\gth$ and $g_\go$, involve $\Z{2}\times\Z{2}$ 
rotations, denoted by $\theta$ and $\omega$, possibly combined with some 
translations. When this is the case such elements are referred to as 
{\it roto--translations}. The phases of the rotations acting on 
$\mathbbm{C}^3$ are 
\equ{
v_\gth = \left(0, \frac{1}{2},-\frac{1}{2}\right)~, 
\quad\text{and}\quad 
v_\go = \left(-\frac{1}{2},0, \frac{1}{2}\right)~,
\label{twists} 
}
respectively.

The action of the space group elements is subsequently extended to the 
left--moving sector of the heterotic worldsheet theory that describes the 
target space gauge degrees of freedom. In a bosonic formulation this sector can 
be described by 16 left--moving coordinates $X_L^I$ ($I=1,\ldots,16$) living on 
a torus $\Real^{16}/\gL_{\E{8}\times\E{8}'}$ defined by the $\E{8}\times\E{8}'$ 
root lattice $\gL_{\E{8}\times\E{8}'}$. The simplest way to extend the 
space group action is the {\it shift embedding} which acts as: 
$g\, X_L^I = X_L^I + 2\pi\, V^I_g$. Hence $V: g \mapsto V_g$ defines a group 
homomorphism of the space group $S$ to the Abelian group $\Real^{16}$ under 
addition. For a general space group element $g = g_\gth^k\, g_\go^l\, g_1^{n_1} 
\cdot\ldots\cdot g_6^{n_6}$, with $k,l=0,1$ and $n_i \in \Intr$, the local 
twist $v_g$ and shift vector $V_g$ can be expanded as  
\equ{ 
v_g = k\, v_\gth + l\, v_\go~, 
\qquad 
V_g = k\, V_\gth + l\, V_\go + n_i\, W_i
\label{VgExpansion} 
}
in terms of the gauge shift vectors $V_\gth$ and $V_\go$ and the discrete 
Wilson lines $W_i$ where summation over $i$ from 1 to 6 is understood. 
In order that $V_g$ defines a proper group homomorphism, it is required that 
\equ{
2\, V_\gth \cong 2\, V_\go \cong 2\, W_i \cong 0~, 
\label{discreteShiftWLs} 
}
where $\cong$ means equal up to $\gL_{\E{8}\times\E{8}'}$ lattice vectors.

The central consistency requirement of heterotic orbifold compactifications is 
modular invariance. For a $\Intr_2\times\Intr_2$ orbifold it requires for all 
commuting space group elements $h, g$ that 
\equ{
V_h \cdot V_{g} - v_h\cdot v_{g} 
 \equiv 0~, 
\label{ModInv} 
} 
where $\equiv$ indicates that both sides are equal up to integers. Combined 
with equation \eqref{discreteShiftWLs} this leads to the following set of 
irreducible modular invariance conditions: 
\equ{
\label{eq:modularinvariance} 
V_\gth^2 \equiv v_\gth^2~, 
\qquad 
V_\go^2 \equiv v_\go^2~, 
\qquad 
V_\gth \cdot V_\go \equiv v_\gth \cdot v_\go~, 
\qquad 
V_\gth \cdot W_i \equiv V_\go \cdot W_i \equiv 0~,
\qquad 
W_i \cdot W_j \equiv 0~,
}
by going through all possible commuting choices of $g,h \in S$.

The spectrum of (twisted or untwisted) closed strings from the sector 
$g \in S$ is dictated by their left-- and right--moving masses
\equ{ 
M_L^2 = \frac 12\, P_\text{sh}^2 + \tN_g - \frac 34~, 
\qquad 
M_R^2 = \frac 12\, p_\text{sh}^2 - \frac 14~, 
\label{MLR2} 
} 
in terms of the (shifted) left-- and right--moving momenta 
\equ{ 
P_\text{sh}=P+V_g~, 
\qquad 
p_\text{sh} = p + v_g~,
\label{Psh}  
}
where $P \in \gL_{\E{8}\times\E{8}'}$ and $p$ is 
from the vectorial or spinorial weight lattice of $\SO{8}$. Here, the twist 
vector $v_g$ is extended to a four--dimensional vector with an extra 0 as first 
component. Furthermore, $\tN_g$ is a (fractional or integer) number operator 
counting the number of left--moving oscillators $\tga_{-n}$ acting on the 
left--moving ground state of the $g$--twisted sector. The physical spectrum is 
subject to the level matching condition $M_L^2 = M_R^2$. The massless states in 
four dimensions have vanishing left-- and right--moving masses, $M_L = M_R = 0$, 
and are subject to the projection conditions
\equ{
V_h \cdot P_\text{sh} - v_h \cdot \left(p_\text{sh} + \Delta\tilde{N}_g\right) \equiv \frac 12 \big( V_g\cdot V_h - v_g\cdot v_h\big)\;, 
\label{Projections}
}
for all space group elements $h$ that commute with $g$, using 
$\Delta\tilde{N}_g^i = \tilde{N}_g^{\bar{i}} - \tilde{N}_g^{i}$, $i=0,1,2,3$, 
where $\tilde{N}_g^{\bar{i}}$ and $\tilde{N}_g^{i}$ are integer oscillator 
numbers counting the numbers of oscillators $\tga_{-n}^{\bar{i}}$ and 
$\tga_{-n}^{i}$ acting on the ground state of the $g$--twisted sector, 
respectively.

\subsection{The standard DW(0--1) $\Intr_2\times\Intr_2$ orbifold}

\begin{figure}[t]
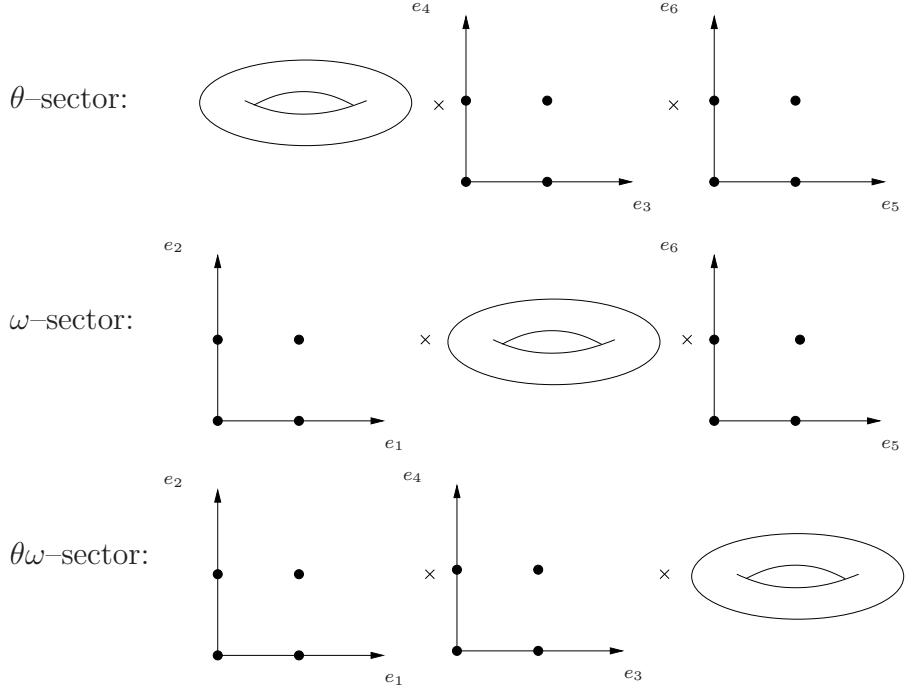

  \centerline{\input FixedTori_DW0-1.pstex_t}\vspace{-0.3cm}
  \caption{Fixed tori of the DW(0--1) orbifold. Every fixed torus 
  intersects 4 + 4 other fixed tori and the intersection loci are points in six dimensions.}
\label{fig:fixedtori0-1}
\end{figure}

Now, we consider the standard $T^6/\Intr_2\times \Intr_2$ orbifold which 
corresponds to the DW(0--1) model of Ref.\ \cite{Donagi:2008xy} in order to see 
how four--dimensional chirality arises. In this case, the elements $g_\gth$ and 
$g_\go$ act only as rotations, hence the space group $S$ is generated by the 
elements: $g_\gth = \big(\theta, 0\big)$, $g_\go =\big(\omega, 0\big)$ and 
$g_i =\big(\Id, e_i\big)$.

When compactifying the heterotic string on this orbifold, massless strings are 
attached to its fixed tori. There are 16 + 16 + 16 fixed tori associated to 
three twisted sectors with orbifold elements $g_\gth$, $g_\go$ and 
$g_\gth g_\go$. These fixed tori are in one--to--one correspondence to the 
space group elements:
\begin{subequations} 
\begin{eqnarray}
\left(\theta,n_i e_i\right)       &\text{for}& n_1 = n_2 = 0 \;\text{and}\; n_3, n_4, n_5, n_6 = 0,1~, 
\\[1ex]
\left(\omega,n_i e_i\right)       &\text{for}& n_3 = n_4 = 0 \;\text{and}\; n_1, n_2, n_5, n_6 = 0,1~, 
\\[1ex]
\left(\theta\omega,n_i e_i\right) &\text{for}& n_5 = n_6 = 0 \;\text{and}\; n_1, n_2, n_3, n_4 = 0,1~, 
\end{eqnarray}
\end{subequations}
and are displayed in figure \ref{fig:fixedtori0-1}. At a given fixed torus 
there exists a six--dimensional $\mathcal{N}=1$ theory (i.e.\ $\mathcal{N}=2$ 
theory in four--dimensional language) with localized hypermultiplets on it. 
Since every fixed torus intersects other fixed tori, six--dimensional 
$\mathcal{N}=1$ supersymmetry is broken to $\mathcal{N}=1$ in four dimensions
at the intersection points. Technically, each fixed torus is orbifolded by the 
action of some other non--trivial elements because the orbifold generators 
$g_\gth$ and $g_\go$ commute. For example, the fixed torus of 
$\left(\theta,0\right)$ is orbifolded by $\left(\omega,0\right)$ and 
$\left(\theta\omega,0\right)$. Hence, the projection conditions 
\eqref{Projections} are active and reduce a hypermultiplet in six dimensions 
to a four--dimensional chiral superfield, giving chiral matter.

For example, in the orbifold standard embedding, where the twists $\theta$ and 
$\omega$ are embedded via the shifts 
$V_\gth = \big(0, \frac{1}{2},\sm\frac{1}{2},0^5\big)\big(0^8\big)$ and 
$V_\go = \big(\sm\frac{1}{2}, 0, \frac{1}{2},0^5 \big)\big(0^8\big)$, 
we obtain a theory with $51$ chiral $\rep{27}$--plets ($3$ untwisted and 
$3\cdot 16$ twisted) and $3$ untwisted chiral $\crep{27}$--plets of $\E{6}$ in 
four dimensions. This precisely corresponds to the hodge numbers of the DW(0--1) 
orbifold: $(h_{11},h_{21})=(51,3)$.

\subsection{The DW(0--2) $\Intr_2\times\Intr_2$ orbifold}
\label{sc:SchoenOrbi}

\begin{figure}[t]
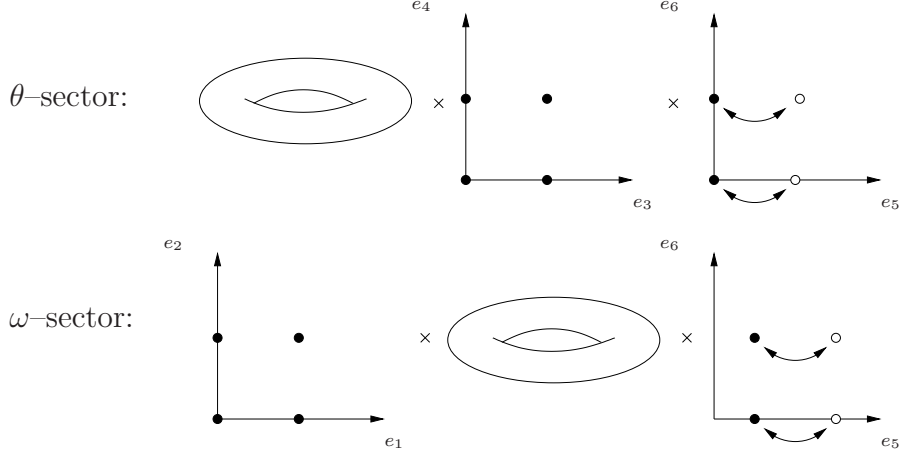

  \centerline{\input FixedTori_DW0-2.pstex_t}\vspace{-0.3cm}
  \caption{Fixed tori of the DW(0--2) orbifold. The fixed tori of the 
  $\theta$--sector never intersect the ones from the $\omega$--sector, as 
  they lie displaced in the third torus.}
\label{fig:fixedtori0-2}
\end{figure}

Next, we turn to the orbifold $T^6/\Intr_2\times \Intr_\text{2,rototrans}$ which 
will be the main focus in this work: the DW(0--2) model in the classification 
\cite{Donagi:2008xy}. Its space group $S$ is generated by the elements 
$g_\gth = \big(\theta, 0\big)$, $g_\go = \big(\omega, \sfrac{1}{2} e_5\big)$ 
and $g_i = \big(\Id, e_i\big)$. 
In detail, the action of $g_\gth$, $g_\go$, $g_\go g_\gth$ and $g_\gth g_\go$ 
on $\mathbbm{C}^3$ is given by
\equ{
\arry{rlcrl}{ 
g_\gth:& \big(z_1,z_2,z_3\big) = \big(z_1,-z_2,-z_3\big)~, 
&\quad & 
g_\go g_\gth: & \big(z_1,z_2,z_3\big) = \big(-z_1, -z_2, z_3 + \sfrac 12\big)~, 
\\[2ex] 
g_\go:& \big(z_1,z_2,z_3\big) = \big(-z_1, z_2, - z_3 + \sfrac 12\big)~, 
& \quad & 
g_\gth g_\go:& \big(z_1,z_2,z_3\big) = \big(-z_1, -z_2, z_3 - \sfrac 12\big)~. 
}
\label{OrbiActions} 
}
This shows explicitly that $g_\gth$ acts as an ordinary $\Intr_2$ rotation, 
while $g_\go$ defines a $\Intr_2$ roto--translation, which we denote by 
$\Intr_\text{2,rototrans}$. The remaining two elements act as an $\Intr_2$ in 
the first two two--tori, but as a translation over half a lattice vector $e_5$ 
in the third two--torus.

This has various important consequences for the distribution of fixed tori 
within the DW(0--2) orbifold: Rather than 16+16+16 there are only 8+8 fixed tori. 
These are left fixed by the group elements $g_\gth$ and $g_\go$. Since 
$g_\gth g_\go$ and $g_\go g_\gth$ act as a pure translations in the third $T^2$, 
they do not produce any fixed tori by themselves, but rather identify the fixed tori 
of $g_\gth$ and $g_\go$ in pairs. The fixed tori of $g_\gth$ and $g_\go$ are 
in one--to--one correspondence to the space group elements, 
\begin{subequations}
\label{FPelements} 
\begin{eqnarray}
g_r = \Big(\theta,n_i e_i\Big) &\text{for}& n_1 = n_2 = 0 \;\text{and}\; n_3, n_4, n_6  = 0,1~, 
\\[1ex] 
g'_{r'} = \Big(\omega,\sfrac{1}{2}e_5 + n_i e_i\Big) &\text{for}& n_3 = n_4 = 0 \;\text{and}\; n_1, n_2, n'_6 = 0,1~. 
\end{eqnarray}
\end{subequations} 
We will often refer to the positions 
$\big(z_1, \sfrac 12 n_3 + \sfrac i2 n_4, \sfrac i2 n_6\big)$ 
and 
$\big(\sfrac 12 n_1 + \sfrac i2 n_2, z_2, \sfrac 14 + \sfrac i2 n_6')$ 
of the fixed tori of $g_\gth$ and $g_\go$ using multi--indices $r=(n_3,n_4,n_6)$ 
and $r'=(n_1,n_2,n_6')$, respectively. As illustrated in figure 
\ref{fig:fixedtori0-2} the fixed tori $r$ and $r'$ lie parallel to each other 
in the third $T^2$. To emphasize this important fact we use the primes on $r'$ 
and $n_6'$ to signal that the 8 fixed points $r'$ are shifted by 
$\frac 14 e_5 = (0,0,\frac 14)$ in the third torus w.r.t.\ the fixed points $r$.

When compactifying the heterotic string on this orbifold, massless strings are 
attached to these fixed tori as for the DW(0--1) orbifold above. However, since 
in this case the fixed tori do not intersect, there are no projections acting 
locally on the six--dimensional $\mathcal{N}=1$ theory. (Or more technically, 
even though $\gth$ and $\go$ commute,the space group elements $g_\gth$ and 
$g_\go$ do not, since they are defined to act on $\Real^6$ not on $T^6$, see
\eqref{OrbiActions}. Hence, the projection condition \eqref{Projections} is not  
implemented for these elements.) Nevertheless, from a four--dimensional 
point--of--view supersymmetry is broken to $\mathcal{N}=1$: There are two 
six--dimensional theories living on the fixed tori of the $g_\gth$ and $g_\go$ 
sectors which realize different six--dimensional $\mathcal{N}=1$ 
supersymmetries. Hence, in the 
effective four--dimensional theory only $\mathcal{N}=1$ remains. As this 
provides an example of non--local supersymmetry breaking, the six--dimensional 
hypermultiplets just branch into two chiral multiplets and consequently the 
resulting four--dimensional theory is necessarily non--chiral. Concretely, for 
the orbifold standard embedding we find for this orbifold $3 + 2\cdot 8 =19$ 
chiral $\rep{27}$--plets and $3+2\cdot 8 = 19$ chiral $\crep{27}$--plets of 
$\E{6}$, i.e.\ the Hodge numbers are $(19,19)$. In fact, because supersymmetry 
is broken non--locally, any DW(0--2) orbifold is non--chiral, independently of 
the choice of shifts and Wilson lines.

Because of this, it would seem that this type of orbifold can never be relevant 
for four--dimensional model building. One of the main messages of this paper is 
that one should not discard such orbifolds for phenomenology just yet. In fact, 
as shown in Section \ref{sc:Models} it is possible to construct an explicit six generation 
$\SU{5}$ GUT model on the resolution of this orbifold. To break the GUT 
to the MSSM and to reduce the number of generations by a factor two one can use 
a (true field--theoretical) Wilson line $W_\text{free}$ associated to a 
free involution $\Intr_\text{2,free}$ of the geometry. In terms of the complex 
coordinates we take this involution to act as 
\equ{
\big(z_1,z_2,z_3\big) \ra \big(z_1 + \sfrac i2, z_2 + \sfrac i2,  z_3 + \sfrac i2\big)~. 
\label{FreeInvolution} 
}
When this involution has been modded out the resulting geometry corresponds to 
the DW(1--3) orbifold with Hodge numbers $(11,11)$ in the classification 
\cite{Donagi:2008xy}. In order that this is a symmetry of the full model, the 
discrete Wilson lines get severely restricted, i.e.\ $W_2\cong W_4\cong W_6$ 
and $W_\text{free} \cong \frac{1}{2} W_2$.

\section{Schoen manifold} 
\label{sc:Schoen} 

The {\it Schoen manifold} $X$ was first introduced in \cite{Schoen:1988}; here 
we follow the description of this manifold given e.g.\ in 
\cite{Bouchard:2007mf,Donagi:2004ub,Donagi:2000zf,Donagi:2000fw}. The Schoen 
manifold is defined as the fiber product 
\equ{
\label{eqn:SchoenFiberProduct}
X = B \times_{\mathbbm{P}^1} B'~, 
}
of two (four--dimensional) rational elliptic surfaces $B$ and $B'$. Hence, the 
manifold $X$ is naturally equipped with two projections $\gp':~ X \ra B$ and 
$\gp:~ X \ra B'$ that project on either factor of the fiber product. Such a 
{\it rational elliptic surface} $B$ is defined as a two--torus fibration 
$\gb: B \ra \mathbbm{P}^1$ over the base $\mathbbm{P}^1$. In terms of the 
fibrations $\gb$ and $\gb'$ the fiber product \eqref{eqn:SchoenFiberProduct} is written as 
\equ{ 
X := \big\{ (p,p') \in B \times B' ~|~ \gb(p) = \gb'(p')   \big\}~. 
}

A two--torus can be described as an elliptic curve, i.e.\ via the 
Weierstrass mapping as the solution to an homogeneous cubic polynomial 
constraint $f(x) = 0$ in the homogeneous coordinates $x=(x_0,x_1,x_2) \in 
\mathbbm{P}^2$. The complex structure of the torus is encoded in the constraint 
$f(x)=0$: A homogeneous cubic polynomial is characterized by $3+6+1=10$ complex 
parameters, of which eight can be removed by complexified $\SU{3}$ rotations of 
$x$ and the overall complex scale is irrelevant. Because of the fibration the 
complex structure in general varies over the base $\mathbbm{P}^1$. Therefore, 
the rational elliptic surface $B$ can be given by
\equ{
B = \arry{c}{\mathbbm{P}^2 \\ \mathbbm{P}^1}\left[ \arry{c}{3 \\ 1} \right]~:
\qquad 
B = \big\{ p=(x, t) \in (\mathbbm{P}^2, \mathbbm{P}^1) ~|~ t_0\, f_0(x) - t_1\, f_1(x) = 0 \big\}~. 
}
The surface $B$ can thus be considered as the blow--up of $\mathbbm{P}^2$ at 
(generically) $3\cdot 3=9$ points $x$ where both cubic polynomials vanish 
simultaneously $f_0(x) = f_1(x) = 0$. Since at each of these points an 
exceptional cycle $\mathbbm{P}^1$ is inserted, the cohomology group 
$H^2(B, \Intr) = \Intr^{10}$ is spanned by the hyper plane class $\ell$ of 
$\mathbbm{P}^2$ and the exceptional classes $e_\rho$ ($\rho=1,\ldots, 9$) with 
intersection numbers $\ell^2 = - e_\rho^2 = 1$. This surface has 
$c_1(B) = 3\, \ell - \sum_\rho e_\rho$ and Euler number 
$\gch(B) = c_2(B) = \gch(\mathbbm{P}^2) + 9\cdot \gch(\mathbbm{P}^1) = 12$.

As the description of $B'$ is similar, the manifold $X$ can be written as a 
complete intersection Calabi--Yau 
\equ{ 
X = \arry{c}{\mathbbm{P}^2 \\ \mathbbm{P}^2 \\ \mathbbm{P}^1}\left[ \arry{cc}{3 & 0 \\ 0 & 3 \\ 1&1} \right]~:
\qquad 
\left\{ 
\arry{c}{
t_0\, f_0(x) - t_1\, f_1(x) = 0 
\\[2ex] 
t_0\, f'_0(x') - t_1\, f'_1(x') = 0 
}~. \right.  
}
Clearly, the Calabi--Yau condition is satisfied as the horizontal sums of the degrees of 
the homogeneous polynomials are one higher than the dimension of the respective 
projective spaces. In this representation the number of complex structure 
deformations is readily counted: There are 4 cubic polynomials, 
$f_0(x), f_1(x), f'_0(x'), f'_1(x')$, each containing 10 complex parameters. 
By redefinitions of $x$, $x'$ and $t$ one can remove $2\cdot 8 +3$ of them and 
two overall complex scales are irrelevant, hence 
$h_{12} = 4\cdot 10 - 2\cdot 8 - 3 - 2 =19$. On both $B$ and $B'$ there are 10 
$(1,1)$--forms corresponding to the classes $\ell, e_\rho$ and $\ell', e_\rho'$, 
respectively. However, since $\gb(p) = \gb'(p')$ there is one linear relation 
among them. Consequently, the number of K\"ahler deformation equals 
$h_{11} = 2 \cdot 10 -1 = 19$. In summary the Hodge numbers of $X$ are $(19,19)$.

\subsection{Singular Schoen manifold: the DW(0--2) orbifold}
\label{sc:SingularSchoen}

Putting the discussion above and the information obtained in Subsection 
\ref{sc:SchoenOrbi} together suggests that the 
$T^6/\Intr_2\times \Intr_{2,rototrans}$ orbifold can be considered as a specific 
singular limit of the Schoen manifold, since their Hodge numbers agree. Indeed, 
by the following considerations this can be confirmed \cite{Donagi:2008xy}: On 
the covering six--torus coordinates $z$ the projects $\gp'$ and $\gp$ act as 
$\gp'(z) = (z_1, z_3)$ and $\gp(z) = (z_2,z_3)$, hence the rational elliptic 
surfaces $B$ and $B'$ are given in this singular limit as 
\begin{subequations} 
\label{EllipticSurfaces} 
\equa{
B &= \big\{ (z_1, z_3) \in T_1^2 \times T_3^2 ~|~ 
(z_1,z_3) \sim (z_1, -z_3) \sim (-z_1, \sfrac 12-z_3) \sim (-z_1,z_3+\sfrac 12) 
\big\}~, 
\\[2ex] 
B' &= \big\{ (z'_2, z'_3) \in T_2^2 \times T_3^2 ~|~ 
(z'_2,z'_3) \sim (z'_2, \sfrac 12-z'_3) \sim (-z'_2,- z'_3) \sim  ( -z'_2, z'_3 +\sfrac 12) 
\big\}~. 
}
\end{subequations} 
These spaces are isomorphic: By applying a change of coordinates 
$(z'_2, z'_3) = (z_1, z_3+\sfrac 14)$ the identifications in $B$ become 
identical to those in $B'$. 
The $\mathbbm{P}^1$ in the fiber product, which has 
the topology of a two--sphere, becomes a rectangular pillow 
\equ{ 
\mathbbm{P}^1 = \big\{ (z_3) \in T_3^2  ~|~ (z_3) \sim (-z_3) \sim (\sfrac 12 - z_3)  \big\}~. 
}

\subsection{Schoen manifold as the resolution of the DW(0--2) orbifold} 
\label{sc:Intersectionnumbers}

As has been explained e.g.\ in \cite{Denef:2004dm,Lust:2006zh,Blaszczyk:2010db} 
one can construct the smooth resolutions of compact orbifolds in a systematic 
fashion. In particular, it is possible to determine a convenient basis of 
divisors and the intersection numbers of the resolution of the orbifold 
$T^6/\Intr_2\times \Intr_\text{2,rototrans}$. Hence, when applying these 
methods to the DW(0--2) orbifold one obtains a smooth Calabi--Yau space which 
constitutes a different realization of the Schoen manifold. As we will use this 
description of the Schoen manifold in the remainder of this paper, we describe 
this procedure in some detail below.

\subsubsection*{Intersection ring of a basis of divisors}

The {\it inherited divisors} $R_a$ ($a=1,2,3$) correspond to the torus divisors 
$\{z_a = c_{a}\}$ with $c_{a}$ some generic complex numbers that are made 
compatible with the orbifold action:
\equ{ 
\arry{c}{ \dsp 
R_1 := \{z_1 = c_{1}\} \cup \{z_1 = -c_{1}\}~, 
\qquad 
R_2  := \{z_2 = c_{2}\} \cup \{z_2 = -c_2\}~, 
\\[2ex] \dsp 
R_3  := \{z_3 = c_3\} \cup \{z_3 =  -c_3\} \cup \{z_3 = \sfrac 12 + c_3\} \cup \{z_1 = \sfrac12 -c_1\}~, 
}
}
Next we define the {\it ordinary divisors} associated to the local coordinates 
defined near the fixed points of the orbifold 
\equ{
\arry{lcl}{\dsp 
D_{1,n_1n_2} := \{ z_1 = \sfrac 12\, n_1+\sfrac i2\, n_2\}~, 
&\quad&  
D_{2,n_3n_4} := \{ z_2 = \sfrac12\, n_3+\sfrac i2\,n_4\}~, 
\\[2ex] \dsp  
D_{3,n_6} := 
\{z_3 = \sfrac i2\,n_6\} \cup \{z_3 =\sfrac 12+\sfrac i2\,n_6\}~, 
&\quad& 
D'_{3,n_6'} := 
\{z_3 = \sfrac 14+ \sfrac i2\,n_6'\} \cup 
\{z_3 = \sfrac 34+ \sfrac i2\,n_6'\}~. 
}
}
Finally, the {\it exceptional divisors}, denoted by $E'_{r'}$ and $E_{r}$, 
arise when we resolve the $\Intr_2$ singularities using e.g.\ toric geometry 
techniques. Since the fixed tori are identified in pairs in the DW(0--2) 
orbifold, we can represent them in terms of the exceptional divisors in the 
fundamental domain of the original $T^6$ (before modding out this 
identification):
\equ{ 
E_r = E_{n_3n_4n_6} = E_{n_3n_40n_6} \cup E_{n_3n_41n_6}~, 
\qquad 
E'_{r'} = E'_{n_1n_2n_6'} = E_{n_1n_20n_6'} \cup E_{n_1n_21n_6'}~. 
} 
These 8+8 exceptional divisors stem from the 8+8 fixed tori displayed 
in the Figure \ref{fig:fixedtori0-2}.

Between these divisors the following linear equivalence relations hold: 
\equ{ 
\arry{c}{\dsp 
2\, D_{1,n_1n_2} = R_1 - \sum_{n_6'} E'_{n_1n_2n_6'}~, 
\qquad 
2\, D'_{3,n_6'} = R_3 - \sum_{n_1,n_2} E'_{n_1n_2n_6'}~, 
\\[3ex] \dsp 
2\, D_{2,n_3n_4} = R_2 - \sum_{n_6} E_{n_3n_4n_6}~, 
\qquad 
2\, D_{3,n_6} = R_3 - \sum_{n_3,n_4} E_{n_3n_3n_6}~.  
}
\label{LinEquivs} 
} 
These linear equivalence relations show that a basis for the $H_2(X,\Real)$ are 
formed by the divisors $R$ and $E$. (In total we have $3+ 2\cdot 8 = 19$ of 
them.) In this basis the Schoen manifold has the following non--vanishing 
self--intersections between these divisors: 
\equ{ 
R_1R_2R_3 = 4~, 
\qquad 
R_2(E'_{n_1n_2n_6'})^2 = R_1(E_{n_3n_4n_6})^2 = -4~. 
} 
Using the linear equivalence relations \eqref{LinEquivs} (self--)intersections 
between any combination of $R$'s, $D$'s and $E$'s are readily computed.

\subsubsection*{Chern classes}

The total Chern class $\text{c}(TX)$ of the tangent bundle of the resolution 
space $X$ can be computed from the splitting principle as 
\equ{
\text{c}(TX) = \prod (1 + D) \, \prod (1+E) \, \prod (1 - R)^2~, 
}
where the products are taken over all divisors of the appropriate types. By 
expanding this out, we can determine the first and second Chern classes of $X$. 
Using the linear equivalence relations \eqref{LinEquivs} we find that 
$\text{c}_1(TX)=0$, confirming that $X$ defines a Calabi--Yau space. For the second Chern 
class we obtain 
\equ{ 
\text{c}_2(TX) = 
- \frac 34 \Big( 
\sum_{r'} (E^\prime_{r'})^2 
+
\sum_{r} (E_{r})^2 
\Big)+\ldots
\label{2ndChern} 
}
The dots $\ldots$ refer to further terms that appear in this expansion in 
principle, but which never contribute when integrated over any four--cycle 
using the intersection numbers given above.

By employing the adjunction formula, $\text{c}_2(D) = D \, \text{c}_2(X|D)$ one 
can determine the Euler number of the hyper surface associated to the divisor 
$D$. In particular, from 
\equ{ 
\gch(R_1) = \text{c}_2(R_1) 
= 24~, 
\quad 
\gch(R_2) = \text{c}_2(R_2) 
= 24~,  
\quad \text{and} \quad
\gch(R_3) = \text{c}_2(R_3) = 0~,
\label{Chern2Ra}
} 
we infer that the divisors $R_1$ and $R_2$ are $K3$ surfaces and $R_3$ is a 
four--torus. Finally, one may consider $D_{1,00}$ and $D_{2,00}$ as the divisor 
classes associated to the rational elliptic surfaces $B'$ and $B$, respectively. 
Indeed, the Euler number of $D_{1,00}$ equals $\gch(D_{1,00}) = 12$, and, since 
$D_{1,00} E_r^2 = -2$, $D_{1,00}$ contains the same fixed points of $g_\gth$ as 
$B'$, see \eqref{EllipticSurfaces}. Hence, we may identify $B'=D_{1,00}$ and 
similarly $B=D_{2,00}$.

\section{Line bundle models on the Schoen resolution}
\label{sc:BundlesOnSchoen}

In this Section we consider Abelian gauge backgrounds on the Schoen geometry 
as described in the previous Section. After that we determine the charged chiral 
spectrum in the presence of this background, showing in particular that even 
with line bundles it is possible to obtain chirality in four dimensions. 
Subsection \ref{sc:BlowupModels} explains how the spectra of such line bundle 
backgrounds can be interpreted as heterotic orbifolds with appropriate blow--up 
modes switched on. The final Subsection illustrates various aspects by giving 
an explicit line bundle model on the Schoen manifold.

\subsection{Abelian gauge flux backgrounds}
\label{sc:GaugeFlux}

On the space $X$ we consider an Abelian gauge background $\cF$ which is embedded 
in the Cartan subalgebra, spanned by the generators $H_I$, of the 
$\E{8}\times\E{8}'$ gauge group of the heterotic theory. In general this gauge 
flux is supported on both the exceptional and inherited divisors 
\equ{
\frac{\cF}{2\gp} = 
\sum_a R_a \, H_{B_a} 
+ \sum_{r} E_{r}\,  H_{V_r}
+  \sum_{r'} E^\prime_{r'}\, H_{V'_{r'}}~.
\label{FullFlux}
}
The {\it line bundle vectors} $V_r, V'_{r'}$ and the magnetic fluxes $B_a$ on 
the tori are sixteen--dimensional component vectors that characterize the 
corresponding line bundle embedding in $\E{8}\times\E{8}'$. To shorten the 
notation we have written $H_A = A_I \, H_I$, where $A$ are referred to as 
bundle or flux vectors with 16 components, $A_I$. 

In order that this gauge background is compatible with the freely acting 
$\Intr_\text{2,free}$ involution \eqref{FreeInvolution} of the orbifold, we 
need to require that 
\equ{
V_{n_3\,n_4\,n_6}= V_{n_3\,n_4+1\,n_6+1}~, 
\qquad 
V'_{n_1\,n_2\,n'_6}= V'_{n_1\,n_2+1\,n'_6+1}~, 
\label{FluxInvolution}
}
Given that the indices take values $n_i = 0,1$, the addition of indices is 
performed modulo 2.

\subsubsection*{Flux quantization}

For a gauge flux configuration \eqref{FullFlux} to be physically admissible it 
has to be integrally quantized, i.e.\ 
\equ{ 
\int_C \frac {\cF}{2\gp} \in \Intr~, 
}
for all curves $C$ that the manifold $X$ admits. This gives a stringent set of 
conditions on the bundle vectors 
\begin{subequations} 
\label{Quant} 
\equ{ 
2\, V'_{n_1n_2n_6'} \cong 2\, V_{n_3n_4n_6} \cong 0~, 
\qquad  
2\, B_1 \cong 2\, B_2 \cong 2\, B_3 \cong 0~,
\label{QuantA}
\\[1ex]
\sum_{n_1,n_2} V'_{n_1n_2n_6'} + B_1\cong 
\sum_{n_3,n_4} V_{n_3n_4n_6} + B_2 \cong 0~, 
\qquad 
 \sum_{n_6'} V'_{n_1n_2n_6'} + \sum_{n_6} V_{n_3n_4n_6} + B_3 \cong 0~.  
\label{QuantB} 
} 
\end{subequations} 
The first two conditions of \eqref{QuantA} are obtained by integrating over the 
curves $D_{2,n_3n_4} E'_{n_1n_2n_6'}$ and $D_{1,n_1n_2} E_{n_3n_4n_6}$, 
respectively. The first two sum conditions in \eqref{QuantB} result from 
integrating over $D_{2,n_3n_4}D'_{3,n_6'}$ and $D_{1,n_1n_2}D_{3,n_6}$, 
respectively. The last sum relation in \eqref{QuantB} is found by integrating 
over $D_{1,n_1n_2}D_{2,n_3n_4}$. When we combine these sum equations with the 
first two conditions of \eqref{QuantA}, the latter three of \eqref{QuantA} are 
inferred. These equations are quite tricky to be solved in general. However, 
there are two (related) ans\"atze that simplify the problem considerably:

First of all, one may assume that the various bundle vectors are either equal 
or opposite, e.g.\ 
\equ{
V_{n_3n_4n_6} = \sum_{s=0,1} (-)^{s(n_4+n_6)}\, V_{s n_3}~, 
\qquad 
V'_{n_1n_2n_6'} = \sum_{s'=0,1}(-)^{s'(n_2+n_6')}\, V'_{s' n_1}~. 
\label{alternateVs} 
}
This particular choice is compatible with the requirement \eqref{FluxInvolution} 
that the gauge fluxes admit the $\Intr_\text{2,free}$ action of equation 
\eqref{FreeInvolution}. In this case, the sums of $V_r$ and $V'_{r'}$ in the 
conditions \eqref{QuantB} form lattice vectors. Consequently, the magnetic 
fluxes of the tori $B_a$ have to be lattice vectors themselves. 

Secondly, taking inspiration from the expansion \eqref{VgExpansion} of the 
local shift vectors $V_g$ in the orbifold construction, one may write the local 
bundle vectors as 
\equ{
\arry{rcl}{
V_{n_3n_4n_6} &=& V_\gth + n_3\, W_3 + n_4\, W_4 + n_6\,W_6 + L_{n_3n_4n_6}~, 
\\[2ex] 
V'_{n_1n_2n_6'} &=& V_\go + n_1\, W_1 + n_2\, W_2 + n'_6\,W_6 + L'_{n_1n_2n'_6}~, 
}
\label{WLinspired}
}
where $2V_\gth \cong 2V_\go \cong 2 W_i \cong 0$ and $L_{n_3n_4n_6} \cong 
L'_{n_1n_2n'_6} \cong 0$. Then, again, the sum conditions \eqref{QuantB} imply 
that the magnetic fluxes $B_a$ are lattice vectors.

\subsubsection*{Bianchi identities}

The central consistency requirements for smooth compactifications are the 
integrated Bianchi identities. In this work we ignore the possibility of 
five--branes, therefore the integrated Bianchi identities have to vanish for 
all divisors $D$, i.e.\
\equ{
\int_D \Big( \tr\cF^2 - \tr \cR^2 \Big) = 0~.
}
Using the expression for the second Chern class \eqref{2ndChern} of $X$, we 
find in the present case that these conditions amount to 
\begin{subequations} 
\label{BIs} 
\equ{ 
B_1 \cdot V_r = 0~, 
\qquad 
B_2 \cdot V'_{r'} = 0~, 
\qquad 
B_1 \cdot B_2 = 0~, 
\label{InnerBIs}
\\[2ex] 
\sum_r (V_r)^2 = 12 + 2\, B_2\cdot B_3~, 
\qquad 
\sum_{r'} (V'_{r'})^2 = 12 + 2\, B_1\cdot B_3~, 
\label{K3likeBIs} 
}
\end{subequations} 
by integrating over $E_r$, $E_{r'}$, $R_3$, and $R_1$, $R_2$, respectively.

The equations in \eqref{InnerBIs} show that the gauge fluxes $B_1$ and $B_2$ 
have to be perpendicular, and, that the gauge flux $B_1$ on $R_1$ is 
perpendicular to all the line bundle vectors $V_r$. Similarly, the gauge flux 
$B_2$ is perpendicular to all vectors $V'_{r'}$. However, there are no 
conditions on the inner products $B_1 \cdot V_{r'}'$ and $B_2 \cdot V_r$. 
The Bianchi identities on the second line, \eqref{K3likeBIs}, are reminiscent 
of the Bianchi identity on a single $\Cplx^2/\Intr_2$ resolution, 
i.e.\ $V^2 = 3/2$ (see e.g.\ Refs.\ \cite{Nibbelink:2007rd,Nibbelink:2007pn}). 
For example, when $B_2$ or $B_3$ vanish and all eight $V_r$ are equal, this 
condition is reproduced identically. The magnetized tori thus lead to 
modifications of the standard local Bianchi identity of the local 
$\Cplx^2/\Intr_2$ resolution.

\subsubsection*{DUY equations and the blow--down limit}

Using that the volume of a divisor $D$ is defined as $\text{Vol}(D) = \int_D J^2/2$, 
the tree--level DUY equation $\int J^2 {\mathcal{F}} = 0$ can be represented as 
\equ{
\sum_a \text{Vol}(R_a) \,{B_a}  
+ \sum_{r} \text{Vol}(E_{r}) \, {V_{r}} 
+ \sum_{r'} \text{Vol}(E_{r'}') \, {V_{r'}'} = 0 ~. 
\label{DUY} 
}
These conditions can be very restrictive. They give an equal number of 
relations between the volumes as the number of linear independent vectors the 
$B_a$, $V_r$ and $V'_{r'}$ can be decomposed in. However, one such relation may 
force many volumes to zero simultaneously, because these volumes are of course 
assumed to be non--negative.

As we describe the Schoen manifold as a resolution of the 
$T^6/\Intr_2\times\Intr_\text{2,rototrans}$, we would like to determine the 
requirements under which a gauge flux configuration allows for a {\it regular 
blow--down limit} in which the underlying six torus $T^6$ has a finite volume. 
Hence, we search for solutions which allow for a full blow-down to the singular 
orbifold, i.e. with $\text{Vol}(E_{r}) =  \text{Vol}(E_{r'}') = 0$ and 
$\text{Vol}(R_a) > 0$. In this case, the DUY equations simplify to 
\equ{
\text{Vol}(R_1) \, B_1 + \text{Vol}(R_2) \, B_2 + \text{Vol}(R_3) \, B_3 = 0~.  
\label{BlowDownDUY}
}
It follows that unless the $B_a$ are linearly dependent or all zero, at least 
some of the volumes $\text{Vol}(R_a)$ are forced to vanish. In particular, if 
only one $B_a$ is non--zero, then the corresponding volume has to be zero 
in the blow--down limit, and hence a regular blow--down limit does not exist. 
Even when $B_3$ is a linear combination of $B_1$ and $B_2$ 
but one of the coefficients is positive, two volumes are forced to zero. Hence, 
only if $B_3$ is a linear combination with negative 
coefficients of $B_1$ and $B_2$, a regular blow--down limit exists. Hence, 
possibly the simplest way to realize this has $B_3 = - B_1 -B_2$. We can cast 
this in the form of two equations
\equ{ 
B_1^2\, \text{Vol}(R_1) + B_1\cdot B_3\,    \text{Vol}(R_3) = 0~, 
\qquad 
B_2^2\, \text{Vol}(R_2) + B_2\cdot B_3\,    \text{Vol}(R_3) = 0~, 
}
since $B_1$ and $B_2$ are perpendicular, see \eqref{InnerBIs}. Since we know 
from \eqref{QuantA} that $2 B_a \cong 0$, we see that the ratios of the volumes 
of the inherited divisors, $\text{Vol}(R_a)/\text{Vol}(R_3)$, $a=1,2$, are fractional.

To allow for a full blow--up we need in addition that the fluxes located at the 
exceptional divisors to be chosen such that the volumes of all of them can be 
taken to be positive at the same time. To ensure this it is again convenient to 
choose that the corresponding bundle vectors are alternating, e.g.\ like in 
\eqref{alternateVs}.

It turns out that the combination of the flux quantization, DUY equations and 
the Bianchi identities is extremely restrictive, hence, to obtain semi--realistic 
models, one is often forced to give up the requirement of a regular blow--down 
limit. (In addition, the one--loop correction to the DUY equation 
\cite{Blumenhagen:2005ga} can force some volumes to be non--vanishing.) 
When this limit does not exist, an orbifold interpretation of the model 
is not ruled out: Often it is possible to shrink quite a number of exceptional 
cycles to zero, while keeping the volumes $\text{Vol}(R_a) > 0$. Hence, locally 
near those shrunken cycles a non--compact orbifold analysis is still possible. 

\subsection{Spectra computation}
\label{sc:SpectraComp}

To determine the physical consequence of models build on such orbifold resolutions 
we need to be able to determine the spectrum of massless states. A convenient way 
of computing the spectrum on an orbifold resolution is provided by the multiplicity 
operator introduced e.g.\ in \cite{Nibbelink:2007rd,Nibbelink:2007pn,Nibbelink:2008tv}. 
Using these methods we can determine both the spectra in four dimensions as well 
as on six--dimensional hyper surfaces.

\subsubsection*{Four--dimensional spectrum}

The spectrum in four dimensions is of key interest in phenomenological studies. 
It can be determined by letting the operator  
\equ{ 
N_{4D}(X) = \int_X \Big\{ 
\frac 16 \, \Big( \frac{\cF}{2\gp} \Big)^3 
+\frac 1{12}\, \text{c}_2(TX) \, \frac{\cF}{2\gp}
\Big\}~, 
}
act on the states contained in the ten--dimensional gaugino. This operator is 
normalized such that it counts the number of chiral superfields. Using the 
intersection numbers determined above  this is computed straightforwardly: 
\equ{ 
N_{4D}(X) = 
2\Big( 1 - \sum_r H_{V_r}^2 \Big)  H_{B_1}
 +   2\Big( 1- \sum_{r'} H_{V'_{r'}}^2\Big) H_{B_2}
+ 4\, H_{B_1} H_{B_2} H_{B_3}~. 
\label{N_4D}
}
The multiplicities of the chiral multiplets in four dimensions are then 
determined by evaluating this operator on the roots of $\E{8}\times\E{8}'$.

There is some tension between solving the Bianchi identities and chirality, 
because of the orthogonality relations, \eqref{InnerBIs}, among the fluxes 
$B_a$, $V_r$ and $V'_{r'}$. However, say $B_1\cdot V_r = 0$, does not imply 
that on all $\text{E}_8\times\text{E}_8'$ roots $H_{B_1} H_{V_r}^2$ vanishes. 
As we show by some examples discussed in the Section \ref{sc:Models}, it is 
indeed possible to obtain a chiral spectrum in four dimensions.

\subsubsection*{Six--dimensional spectra on divisors}

In addition, we can define the multiplicity operator $N_{6D}(D)$ in six 
dimensions for any divisor $D\subset X$. Positive values of $N_{6D}$ count the 
number of half--hyper multiplets, while negative values count (two times) the 
number of vector multiplets. (In six dimensions the fermions of hyper and vector 
multiplets have opposite chirality.) As integrals over the whole space $X$ 
they read 
\equ{ 
N_{6D}(D) = 
\int_X D\, \Big\{ 
\frac 12\, \Big( \frac{\cF}{2\gp} \Big)^2 + \frac 1{12}\, \text{c}_2(D)
\Big\}~, 
} 
where $\text{c}_2(D)$ are given in \eqref{Chern2Ra}. Given that the divisors 
$R_1,R_2$ may be interpreted as $K3$ surfaces and $R_3$ as a four--torus, the 
spectra on these divisors are probably the most interesting. Using the 
intersection numbers we readily compute this explicitly for $D= R_a$: 
\begin{subequations} 
\label{N_6D}
\equa{
N_{6D}(R_1) &= 2\Big( 1 - \sum_r H_{V_r}^2 \Big) 
+ 4\, H_{B_2}H_{B_3}~, 
\label{N_6DR1} 
\\[1ex] 
N_{6D}(R_2) &= 2\Big( 1- \sum_{r'} H_{V'_{r'}}^2\Big) 
+ 4\, H_{B_1}H_{B_3}~, 
\\[1ex] 
N_{6D}(R_3) &= 4\, H_{B_1} H_{B_2}~. 
}
\end{subequations} 

\subsubsection*{Relation between the six-- and four--dimensional spectra}

As explained in Subsection \ref{sc:SchoenOrbi} the orbifold 
$T^6/\Intr_2\times\Intr_\text{2,rototrans}$ never leads to four--dimensional 
chirality. The reason is basically that such models only contain hypermultiplets in 
six dimension, which simply branch to vector--like combinations of chiral 
multiplets in four dimensions. In the smooth case we have found a way to 
bypass this no--go. The key here are the magnetic fluxes $B_a$ on the 
divisors $R_a$ that correspond to the tori of the orbifold in the blow--down 
limit. Indeed, if we set all $B_a =0$, then \eqref{N_4D} simply says that 
$N_{4D} = 0$: no chirality. Hence, precisely by allowing for magnetized 
divisors $R_a$ we can avoid this no--go and obtain chirality.

To see that this effect is expected from field theory, let us consider the case 
in which the flux $B_2$ has been switched off. The four--dimensional multiplicity 
operator \eqref{N_4D} then leads to a relation between the six--dimensional 
spectrum on $R_1$ given by \eqref{N_6DR1} and the spectrum in four dimensions: 
\equ{
N_{4D}(X) = H_{B_1}\, N_{6D}(R_1)~.  
\label{Nfactorization}
} 
This equation can be interpreted as follows: When we compactify on a 
$K3 \times T^2$, then we can consider first the six--dimensional 
theory that results from the compactification on $K3$. This effective 
six--dimensional model is subsequently reduced on a two--torus. 
It is well--known that if there is no magnetic flux, this second step 
results in a vector--like spectrum in four dimensions. However, if there is a 
magnetic flux $B$ present, a chiral spectrum arises: only the chiral 
fermionic states of charge $q$ for which $Bq > 0$ survive, and their 
multiplicity is given by $Bq$ provided that the smallest charge in the 
spectrum is unity \cite{Cremades:2004wa,Abe:2009uz}. It is intriguing to 
notice that \eqref{Nfactorization} says exactly this: $N_{6D}(R_1)$ determines 
the spectrum in a six--dimensional world. The operator $H_{B_1}$ gives the 
charge $B_1\cdot w$ of the six--dimensional state associated with the 
$\E{8}\times\E{8}'$ root $w$ under the magnetic flux $B_1$.

When all three fluxes are switched on, the relation between the four-- and 
six--dimensional spectra apparently reads
\equ{
N_{4D}(X) = H_{B_1}\, N_{6D}(R_1) + H_{B_2}\, N_{6D}(R_2)  - H_{B_3}\, N_{6D}(R_3)~.
}
This follows directly by identifying the six--dimensional multiplicity 
operators \eqref{N_6D} in the four--dimensional expression \eqref{N_4D}. The 
final term corrects for over counting of states charged under $B_1$, $B_2$ and 
$B_3$ simultaneously.

\subsection{Interpretation as blow--up of DW(0--2) orbifold models}
\label{sc:BlowupModels}

So far, we have analyzed the Schoen geometry as a smooth Calabi--Yau 
and described line bundle backgrounds on it. The fact that the Schoen manifold 
is the resolution of the DW(0--2) orbifold, as discussed in Section \ref{sc:Schoen}, 
has essentially been irrelevant in our investigation. Now, we would like to 
describe how a given line bundle model can be understood as a heterotic DW(0--2) 
orbifold model with a certain number of blow--up modes attaining vacuum 
expectation values (VEVs). We first recall how this analysis can be done in 
general using a $\mathcal{N}=1$ language in four dimensions following 
\cite{Nibbelink:2008tv,Nibbelink:2009sp,Blaszczyk:2010db}. After that we 
conclude this Subsection by describing this procedure in a six--dimensional 
supersymmetric formulation which is more appropriate since the DW(0--2) orbifold 
model has $\mathcal{N}=1$ supersymmetric sectors in six dimensions.

\subsubsection*{Four dimensional $\boldsymbol{\mathcal{N}=1}$ language}

In heterotic orbifolds a fixed point gets blown up if a twisted chiral superfield
$\Phi_\text{bm}^{(r)}$, localized at that fixed point, acquires a 
non--vanishing VEV: $\langle \Phi_\text{bm}^{(r)} \rangle \neq 0$. 
The value of this VEV determines the volume of the exceptional cycle $E_r$ that 
appears in this resolution process. As we recalled in Section 
\ref{sc:GeneralFeatures} any twisted state is characterized by a shifted 
left--moving momentum $P_\text{sh}$. In Refs.\ \cite{Nibbelink:2009sp,Blaszczyk:2010db,Nibbelink:2010wm} 
it was realized that, as long as this twisted state does not involve any 
oscillator excitations, its shifted momentum $P_\text{sh}$ precisely determines the 
local line bundle vector $V_r$ associated to the exceptional divisor $E_r$. 
Some special cases might occur: Sometimes it happens that a bundle vector 
corresponds to a blow--up mode that has been projected out by the orbifold 
action in the four--dimensional theory. It is also possible that the bundle 
vector is associated to a massive state in the orbifold spectrum. 

The spectrum of the orbifold model and the one of the blow--up theory are 
generically not identical, but closely related: First of all the VEVs of the 
twisted states $\langle \Phi_\text{bm}^{(r)} \rangle$ lead to some gauge symmetry 
breaking. Furthermore, the blow--up modes are not present in the blow--up 
spectrum as charged states, but rather as (complexified) axions $b_r$. The 
relation between the blow--up mode and the axion reads 
\equ{ 
\Phi_\text{bm}^{(r)} = e^{b_r}\, \langle \Phi_\text{bm}^{(r)} \rangle~. 
}
In the smooth description this axion generically gives a mass to the gauge field of the 
broken $\U{1}$ via the Stueckelberg mechanism. In the blow--up picture the 
$\U{1}$ is broken just by a standard Higgs mechanism.

As a consequence of this gauge symmetry breaking the representations of matter 
fields get branched. But still, this is not enough to match the orbifold and 
resolution spectra \cite{Nibbelink:2009sp,Blaszczyk:2010db}: One 
needs to preform field redefinitions of the other twisted matter states 
$\Phi_\text{orb}$ involving the corresponding blow--up modes to obtain an 
exact agreement of the spectra, i.e.\ 
\equ{ 
\Phi_\text{orb} = e^{\pm b_r}\, \Phi_\text{res}~. 
}
The signs $\pm$ have to be chosen appropriately to ensure that the weights of 
$\Phi_\text{res}$ are $\E{8}\times\E{8}'$ roots, while those of $\Phi_\text{orb}$ 
belong to the shifted weight lattice defined in \eqref{Psh}.

\subsubsection*{Blow--ups in six dimensions}

Before we describe the blow--up procedure in a six--dimensional language, we 
first briefly recall some properties of $\mathcal{N}=1$ theories in six 
dimensions. There are three basic irreducible representation of $\mathcal{N}=1$ 
supersymmetry relevant for our discussion: 
i) A vector multiplet $\mathcal{V} = (\text{V}, \Phi)$ contains a vector 
superfield $\text{V}$ and a chiral superfield $\Phi$ from the 4D $\mathcal{N}=1$ 
perspective. 
ii) A hypermultiplet contains two independent chiral superfields 
$\mathcal{H} = (\Phi, \Phi^c)$ that live in charge--conjugate representations. 
This means that the gauge properties of the hypermultiplet is uniquely 
specified by the representation and $\U{1}$ charges of either chiral component. 
iii) Finally, a half--hyper multiplet is a hypermultiplet with a certain 
reality condition imposed. Therefore, it has only half of the number of 
independent components as a normal hypermultiplet. In other words, using the 
four--dimensional $\mathcal{N}=1$ terminology, a half--hyper is a chiral 
superfield in a real or pseudo--real representation.

Now, if a twisted hypermultiplet plays the role of a blow--up mode in order to 
resolve a fixed torus, then only one of its chiral superfield components 
actually takes a VEV, while the other component only gets redefined: 
\equ{ 
\mathcal{H}_\text{bm}^{(r)} = e^{b_r}\, 
(\langle \Phi_\text{bm}^{(r)}\rangle, \Phi^{c(r)}_\text{bm,res})~. 
\label{HyperBM} 
}
Because the chiral superfield components of a hypermultiplet carry opposite 
$\U{1}$ charges, they have to be redefined with opposite powers of the 
blow--up mode: 
\equ{
\mathcal{H}_\text{orb} = (\Phi_\text{orb},\Phi^c_\text{orb}) = 
\big( e^{\pm b_r}\, \Phi_\text{res}, e^{\mp b_r}\, \Phi^c_\text{res} \big)~, 
\label{HyperRedef} 
}
for appropriate choice of sign $\pm$. After these field redefinitions the 
chiral superfields in the blow--up mode hypermultiplet do not seem to fall into 
proper $\mathcal{N}=1$ representations anymore. However, this does not signify 
that the blow--up breaks six--dimensional supersymmetry: The remaining chiral 
superfield components will be completely neutral, and therefore form 
half--hypermultiplets by themselves.

\subsection{Sample model: An eight generation GUT} 
\label{sc:aGUT}

We conclude this Section with a concrete example of a line bundle model which 
is constructed on the Schoen geometry to illustrate many aspects of the general 
description developed in this and the preceding Sections. Consider the following 
particular line bundle model on the resolution of our $T^6/\Intr_2\times \Intr_\text{2,rototrans}$: 
\equ{ 
B_3 = - B_1~, 
\quad B_2 = 0~, 
\qquad 
V_{n_3n_4n_6} = (-)^{n_4 + n_6}\, V_{0}~, 
\quad 
V'_{n_1n_2n_6'}
(-)^{n_2+n_6'}\, V'_{n_1}~,
\label{SampleInput}
}
with 
\equ{
\arry{l}{
B_1 = \big(1,1,1,\sm1,0, 0^3\big)\big(0^8\big)~, 
\\[2ex] 
V_0 = \big( 0, 0, \sfrac 12, \sfrac 12,0, 0^3\big)\big(1,0^7\big)~, 
} 
\qquad
\arry{l}{
V'_0 = \big(0,0,0, \sfrac 12, \sfrac 12, 0^3\big)\big(0^8\big)~, 
\\[2ex] 
V'_1 = \big(\sfrac 12,\sm\sfrac 12, 0, 0,0, 0^3\big)\big(0^8\big)~. 
}
}
This model exhibits the following properties:

The bundle vectors satisfy all the requirements specified in Section 
\ref{sc:GaugeFlux}: The quantization conditions \eqref{Quant} are fulfilled, 
because the alternating signs in the $V_r$ and $V'_{r'}$ are in accordance with 
\eqref{alternateVs} and the vectors $B_a$ are lattice vectors.

They also satisfy all Bianchi identities \eqref{BIs}: $B_1$ is perpendicular 
to all vectors $V_r$. Since all vectors $V_r$ square to $3/2$ and $B_2=0$ the 
first condition in \eqref{K3likeBIs} is satisfied. The second condition in 
\eqref{K3likeBIs} is satisfied as well, since both sides are equal: 
\equ{ 
\sum_{r'} V'_{r'}{}^2 = 8 \cdot \sfrac 12 = 4~, 
\qquad 
12 + 2\, B_1 B_3 = 12 -2\cdot 4 = 4~. 
}
Because $B_3=-B_1$ and $B_2=0$ a blow--down of this model is allowed by the DUY 
equations \eqref{DUY} while keeping the torus radii, set by the volumes of the 
divisors $R_a$, finite. In the blow--down limit, $\text{Vol}(E_r) = 
\text{Vol}(E'_{r'})=0$, the volumes of $R_1$ and $R_3$ have to be equal, 
$\text{Vol}(R_3) = \text{Vol}(R_1)$. The alternating signs of $V_r$ and 
$V'_{r'}$ ensure that the DUY equations also allow for a finite blow--up of all 
exceptional cycles.

The gauge group that is left unbroken by this Abelian gauge configuration is
\equ{
\SU{5} \times \SO{14}' \times \U{1}^5~,
}
from the first and second $\E{8}$ group factor. Since for this choice of bundle 
vectors $B_2=0$ and all $V_r$ are equal up to a sign, the 4D multiplicity 
operator \eqref{N_4D} reduces to 
\equ{ 
N_{4D} = 2\, H_{B_1} \Big( 1 - 8\, H_{V_0}^2 \Big)~.
\label{Mult4DOneTerm}
}
The resulting spectrum,
\equ{ 
8\, (\rep{10}, \rep{1}) + 12\, (\crep{5},\rep{1}) + 4\, (\rep{5}, \rep{1}) + 24\, (\rep{1},\rep{1})~,
} 
is chiral w.r.t.\ to the five $\U{1}$ charges (which we omitted for notational 
simplicity). (W.r.t.\ the hidden gauge group at most a purely vector--like 
spectrum arises, which is invisible for the multiplicity operator.) Hence, the 
model might be considered as an eight generations $\text{SU}(5)$ GUT toy--model 
with four Higgs pairs.

\section{A line bundle MSSM on the Schoen manifold}
\label{sc:Models}

We present an MSSM--like model with three generations as a line bundle model on 
the resolution of $T^6/\Intr_2\times \Intr_\text{2,rototrans}$. In the first 
Subsection we construct an $\SU{5}$ GUT model with six generations on the 
Schoen manifold using line bundles. In Subsection \ref{sc:WfreeMSSM} we identify a 
Wilson line that can be associated with a freely acting involution, which both 
reduces the number of generations to three and breaks the gauge group to the 
standard model group. In the next Subsection we show that a $K3$ subspace of 
the Schoen manifold can be blown down to a four--dimensional orbifold 
$T^4/\Intr_2$ on which the model can be quantized using standard CFT 
techniques. In Subsection \ref{sc:blowupexample} we use this to give an 
alternative description of the line bundle MSSM on the Schoen manifold in terms 
of a blow--up of this orbifold with a magnetized torus.

\subsection{Six GUT generations on the Schoen resolution}
\label{sc:realisticGUT}

We define a line bundle model on the Schoen manifold with the flux vectors
\equ{ 
B_1 = \big(3,-3,0^6\big)\big(3,3,0^6\big) \quad\text{and}\quad B_2 = B_3 = 0~, 
\label{MSSMFlux} 
}
on the ordinary divisors $R_a$,
\begin{subequations}
\label{eqn:MSSMResolutionData1}
\equa{
V_{(0,0,0)}  =  V_{(0,1,0)} = -V_{(0,0,1)} = -V_{(0,1,1)} &= 
\big( \sfrac 14^8                 \big)\big( 0,         0, 0, \sfrac 12, 0,-\sfrac 12,-\sfrac 12,-\sfrac 12\big)~, 
\\[1ex] 
V_{(1,0,0)} =  V_{(1,1,0)} = -V_{(1,0,1)} = -V_{(1,1,1)} &=
\big(    0, \sfrac 12, \sfrac 12, 0^5\big)\big( 0, \sfrac 12, 0,         0, 0,-\sfrac 12,-\sfrac 12,-\sfrac 12\big)~, 
}
\end{subequations}
on the exceptional divisors $E_r$, and finally,
\begin{subequations}
\label{eqn:MSSMResolutionData2}
\begin{eqnarray}
V'_{(0,0,0)} = -V'_{(0,1,1)} &=& 
\big( 0,-\sfrac 12,  -\sfrac 12, 0^5\big)\big( \sfrac 12, \sfrac 12, \sfrac 12, 0,-\sfrac 12, 0, 0, 0\big)~, 
\\[1ex] 
V'_{(0,1,0)} = -V'_{(0,0,1)} &=&   
\big( 0,-\sfrac 12,  -\sfrac 12, 0^5\big)\big( \sfrac 12, \sfrac 12,-\sfrac 12, 0, \sfrac 12, 0, 0, 0\big)~, 
\\[1ex] 
V'_{(1,0,0)} = V'_{(1,1,0)} &=& 
\big( 0,         1,           0, 0^5\big)\big(-\sfrac 12,-\sfrac 12,         0, 0,         0, 0, 0, 0\big)~, 
\\[1ex] 
V'_{(1,1,1)} = V'_{(1,0,1)} &=& 
\big(-1,0^7\big)\big( -\sfrac 12,-\sfrac 12,0^6\big)~,  
\end{eqnarray}
\end{subequations}
on the exceptional divisors $E'_{r'}$.

This choice of bundle vectors fulfills the quantization conditions 
\eqref{Quant} and the DUY equations \eqref{DUY} for appropriately 
chosen volumes. All bundle vectors $V_r$ and $V'_{r'}$ have 
$V_r^2 = V'_{r'}{}^2 = 3/2$. This is consistent with the Bianchi identities 
\eqref{K3likeBIs}, which reduce to 
\equ{ 
\sum_r (V_r)^2 = \sum_{r'} (V'_{r'})^2 = 12~; 
}
since there are no corrections resulting from magnetic fluxes $B_a$ as only 
$B_1 \neq 0$. The unbroken gauge group in this gauge configuration reads
\equ{
\SU{5} \times \SU{5}' \times \U{1}^8~. 
}
The four--dimensional multiplicity operator \eqref{N_4D} is computed 
straightforwardly and the resulting chiral spectrum is given in Table 
\ref{tab:ExampleMagnetizedSpectrum}. In this Table we have distinguished the 
various states, in particular the singlets, by their eight $\U{1}$ charges 
$(q_0,\ldots, q_7)$. Notice that, curiously, this model has six generations 
in both, the observable and the hidden, $\SU{5}$.

\begin{table}[t!]
\begin{center}
\begin{tabular}{|c|c|cccccccc|}
\hline   
Superfield   & Representation        & \multicolumn{8}{|c|}{$\U{1}$ charges} \\                       
multiplicity & $\SU{5}\times\SU{5}'$ & $q_{0}$ & $q_{1}$ & $q_{2}$ & $q_{3}$ & $q_{4}$ & $q_{5}$ & $q_{6}$ & $q_{7}$ \\
\hline\hline
     6  & $\left(\crep{10},  \rep{1}\right)$ & 0 & 0 & 0 & 0 & 1 & 0 &-3 & 0 \\
     6  & $\left(  \rep{5},  \rep{1}\right)$ & 0 & 0 & 0 & 0 & 0 & 0 &-6 & 0 \\
     6  & $\left( \crep{5},  \rep{1}\right)$ & 1 & 0 & 1 & 0 &-1 & 0 & 1 & 0 \\
     6  & $\left(  \rep{5},  \rep{1}\right)$ & 1 & 0 & 1 & 0 & 0 & 0 & 4 & 0 \\ 
    24  & $\left(  \rep{1},  \rep{1}\right)$ & 2 & 0 & 0 & 0 & 0 & 0 & 0 & 0 \\
     6  & $\left(  \rep{1},  \rep{1}\right)$ &-1 & 0 &-1 & 0 &-1 & 0 & 5 & 0 \\
     6  & $\left(  \rep{1},  \rep{1}\right)$ & 1 & 0 &-3 & 0 & 0 & 0 & 0 & 0 \\
     6  & $\left(  \rep{1},  \rep{1}\right)$ & 0 & 0 & 0 & 0 & 2 & 0 & 0 & 0 \\    
 \hline\hline 
     6  & $\left(  \rep{1},\crep{10}\right)$ & 0 & 0 & 0 & 2 & 0 & 0 & 0 &-6 \\
    24  & $\left(  \rep{1},  \rep{5}\right)$ & 0 & 1 & 0 & 3 & 0 & 0 & 0 &-2 \\
     6  & $\left(  \rep{1}, \crep{5}\right)$ & 0 & 0 & 0 &-2 & 0 & 0 & 0 &-8 \\
     6  & $\left(  \rep{1}, \crep{5}\right)$ & 0 & 0 & 0 & 0 & 0 & 1 & 0 & 7 \\
     6  & $\left(  \rep{1}, \crep{5}\right)$ & 0 & 0 & 0 & 0 & 0 &-1 & 0 & 7  \\ 
    42  & $\left(  \rep{1},  \rep{1}\right)$ & 0 & 0 & 0 & 4 & 0 & 1 & 0 &-5 \\
    42  & $\left(  \rep{1},  \rep{1}\right)$ & 0 & 0 & 0 & 4 & 0 &-1 & 0 &-5 \\
    24  & $\left(  \rep{1},  \rep{1}\right)$ & 0 & 1 & 0 &-3 & 0 & 1 & 0 &-5 \\
    24  & $\left(  \rep{1},  \rep{1}\right)$ & 0 & 1 & 0 &-3 & 0 &-1 & 0 &-5 \\
     6  & $\left(  \rep{1},  \rep{1}\right)$ & 0 & 2 & 0 & 0 & 0 & 0 & 0 & 0 \\
\hline
\end{tabular}
\end{center}
\caption{This line bundle model on the Schoen manifold has six generations of 
$\SU{5}$ in both, the observable and the hidden, sectors. States in the 
first block are charged under the observable $\E{8}$; states in the second block 
are charged under the hidden group.}
\label{tab:ExampleMagnetizedSpectrum}
\end{table}

\subsection{Freely acting $\Z{2}$ and MSSM with three generations}
\label{sc:WfreeMSSM}

One can define a freely acting involution $\Z{2,\text{free}}$ as in equation 
\eqref{FreeInvolution} that reduces the number of generations by a factor 
$1/2$. In addition, the freely acting involution can be embedded as a Wilson 
line that breaks $\SU{5}$ to $\SU{3}\times\SU{2}\times\U{1}_Y$. We take this 
Wilson line,  
\begin{equation}
\label{eq:SU5BreakingFreelyActingZ2}
W_\text{free} =  \big(0^3, 1, 1, 1,-\sfrac 32,-\sfrac 32 \big) \big(0^8\big)~,
\end{equation}
to point in the standard hypercharge direction of $\SU{5}$. This choice of 
$W_\text{free}$ fixes the first $\SU{5}$ to define the observable sector and 
leads to an MSSM--like model with three generations.

Contrary to the situation in field theory, there are further requirements on 
this Wilson line in string theory \cite{Blaszczyk:2009in,Blaszczyk:2010db}: It 
has to satisfy $2 W_\text{free} \cong W_2 \cong W_4 \cong W_6 \cong 0$ and it 
has to respect the modular invariance conditions 
\begin{equation}
2W_\text{free}^2 \equiv W_\text{free} \cdot W_i \equiv 0~. 
\end{equation}
These additional conditions were derived in context of orbifold constructions 
where $\Intr_\text{2,free}$ is part of the space group.

\subsection{Singular limits of the Schoen GUT with line bundles}
\label{sc:OrbiLimits}

\subsubsection*{Full blow down limit}

Taking the magnetic flux $B_1$ to vanish for a moment, we can consider the full 
blow--down limit of the GUT model with six generations. It has an exact 
heterotic orbifold CFT as formulated in Section \ref{sc:SchoenOrbi} as the 
$T^6/\Intr_2\times \Intr_\text{2,rototrans}$ orbifold with a definite choice of 
gauge shifts, $V_\gth,V_\go$, and discrete Wilson lines, $W_i$. As dictated by 
the flux vectors \eqref{eqn:MSSMResolutionData1} and \eqref{eqn:MSSMResolutionData2}
they are given by
\begin{equation}
\begin{array}{ll}
 V_\gth = \big( \sfrac 14^8\big) \big( \sfrac 12,-\sfrac 12, \sfrac 12, 0, \sfrac 12, 0^3\big)~,         & V_\go = \big( 0, -\sfrac 12, -\sfrac 12,  0^5\big) \big(  -\sfrac 12, -\sfrac 12, -\sfrac 12,  0,  \sfrac 12, 0^3\big)~,\!\!\!\!\! \label{eq:ShiftsandWilson} \\[2ex]
 W_1 = \big( 0,  \sfrac 12, -\sfrac 12,  1,  1,  0^3\big) \big(   0, 1, \sfrac 12, 1, -\sfrac 12, 0^3\big)~, & W_3 = \big( -\sfrac 14, \sfrac 14, \sfrac 14,  -\sfrac 14^5\big) \big( 0, \sfrac 12, 0, \sfrac 12,-1, 0^3\big)~,
\end{array}
\end{equation}
and the other Wilson lines vanish. This choice fulfills the conditions of 
modular invariance \eqref{eq:modularinvariance}. As discussed in Section 
\ref{sc:SchoenOrbi}, the spectrum of this orbifold can be computed using 
orbifold CFT techniques but is necessarily non--chiral as long as no magnetic 
fluxes $B_a$ have been reintroduced.

\subsubsection*{The $\boldsymbol{T^4/\Z{2}}$ orbifold inside the Schoen manifold}

Since the Schoen model defined in this Section has only a single magnetic flux, 
$B_1$, switched on, see \eqref{MSSMFlux}, the DUY equations \eqref{DUY} imply 
that in a full blow--down the volume of $R_1$ has to vanish as well. However, 
we can exploit that there also exists a partial blow--down in which all 
$\text{Vol}(E_r) \ra 0$ while the volumes of all inherited divisors $R_a$ and 
of at least some other exceptional divisors $E'_{r'}$ stay finite. Therefore, 
this partial blow--down leads to an intermediate $T^4/\Z{2}$ orbifold with 
torus coordinates $(z_2, z_3)$ on which the $\Z{2}$ action acts via the twist 
$v_\gth$ given in \eqref{twists} (c.f.\ \cite{Buchmuller:2012mu}).

For this intermediate $T^4/\Intr_2$ orbifold an exact heterotic CFT description 
exists. Taking its gauge embedding as given by $V_\gth$ and $W_3$ from equation 
\eqref{eq:ShiftsandWilson}, its low energy limit results in a model with 
$\mathcal{N} = 1$ supersymmetry in six dimensions with gauge group
\equ{
\E{6} \times \SU{8}' \times \U{1}^3~. 
}
The spectrum of hypermultiplets including $\U{1}$ charges of this intermediate 
six--dimensional orbifold theory is computed using \cite{Nilles:2011aj} and 
listed in the first column of Table \ref{tab:OrbifoldBlowUp}.

A simple, yet non--trivial crosscheck of this spectrum is that it is free of 
irreducible gravitational anomalies, e.g.\ that the sum condition 
$\#(\text{hyper})-\#(\text{vector}) = 244$ holds. Indeed, using Table 
\ref{tab:OrbifoldBlowUp} it is straightforward to count the number of vector-- and 
hypermultiplets: 
\equ{
\#(\text{vector}) = 78 + 63 + 3\cdot 1 = 144~,
\quad 
\#(\text{hyper})  =  2 \cdot \big(27 + \sfrac 12\cdot 70 + 2\big) + 16 \cdot 2 \cdot 8 +4 = 388~,
}
where the factor $\frac 12$ accounts for the fact that the 
$(\rep{1},\rep{70})_{(0,0,0)}$ is a half--hyper. The last $4$ additional hypers 
correspond to untwisted moduli which are not displayed in Table 
\ref{tab:OrbifoldBlowUp}.

\subsection{Schoen line bundle MSSM as a blown up orbifold}
\label{sc:blowupexample}

The MSSM--like model of Subsection \ref{sc:realisticGUT} can now be reproduced 
as a blow--up of the $T^4/\Intr_2$ orbifold discussed in the Subsection above 
equipped with a magnetic flux on the torus to generate four--dimensional 
chirality. In short, this procedure reads: 
\begin{enumerate}
\item Blow--up the $T^4/\Z{2}$ orbifold to a smooth $K3$ manifold by giving 
      VEVs to 16 blow--up modes and use field redefinitions to obtain the 
      spectrum on $K3$.
\item Turn on the additional fluxes on the divisors $E'_{r'}$, decompose 
      gauge group and branch the representations accordingly.
\item Generate four--dimensional chirality by switching on the magnetic flux 
      $B_1$ as well. 
\end{enumerate}
In this process the magnetic flux $B_1$ does not lead to breaking of four--dimensional 
supersymmetry since the contribution from the fluxes on $E'_{r'}$ cancels the one from 
$B_1$ in the DUY equations \eqref{DUY}.
In the following we describe this procedure in detail:

\subsubsection*{Blowing up the intermediate $\boldsymbol{T^4/\Intr_2}$ orbifold}

The intermediate $T^4/\Z{2}$ orbifold gets blown up to a $K3$ surface by 
assigning VEVs to the blow--up modes, i.e.\ to 16 twisted states localized at 
the 16 singularities of the $T^4/\Z{2}$ orbifold. At each singularity (labeled 
by the multi--index $r=(n_3,n_4,n_5,n_6)$) a blow--up mode, 
$\Phi^{(r)}_\text{bm}$ contained in a twisted hypermultiplet, is chosen such 
that its shifted left--moving momentum $P^{(r)}_\text{sh}$ agrees with the flux 
vector $V_r$ localized on the divisor $E_r$: 
\begin{equation}
P^{(r)}_\text{sh} = V_r  \quad\text{for all }\quad r=(n_3,n_4,n_5,n_6)~.
\end{equation}
All these blow--up modes are chosen to be in eight--dimensional representations of the 
hidden $\SU{8}'$ gauge group of the $T^6/\Intr_2$ orbifold, so that, 
consequently, the gauge group gets broken to
\equ{
\E{6} \times  \SU{7}' \times \U{1}^4~. 
}

As explained in Subsection \ref{sc:BlowupModels}, when the blow--up mode 
$\Phi^{(r)}_\text{bm}$ in a given twisted sector attains a VEV, field 
redefinitions, \eqref{HyperBM} and \eqref{HyperRedef}, have to be performed on 
the other states in the same twisted sector in order to ensure that all fields 
in the blow--up are characterized by $\E{8}\times\E{8}'$ roots. The appropriate 
field redefinitions required by this blow--up procedure are listed in the 
second column of Table \ref{tab:OrbifoldBlowUp}. As the phases of the 
orbifold blow--up modes have been 
reinterpreted as axions, the remaining hypermultiplet components do not seem to 
form proper six--dimensional $\mathcal{N}=1$ hypermultiplets. However, as can 
be verified from this Table, these chiral superfield are neutral and can thus 
be interpreted as half--hypermultiplets.

\subsubsection*{Additional fluxes on the exceptional divisors $\boldsymbol{E'_{r'}}$}

Up to now, we have turned on fluxes only on the exceptional divisors $E_r$, 
which correspond to the blown--up fixed points of the $T^4/\Z{2}$ orbifold. 
After the field redefinition to the blow--up field basis, the charges w.r.t.\ 
to the four $\U{1}$ factors are taken such that they correspond to the first 
four charges $(q_0,q_1,q_2,q_3)$ in Table \ref{tab:ExampleMagnetizedSpectrum}. 
The $\U{1}$'s associated to the charges $(q_0,q_1,q_2)$ already exist at the 
intermediate $T^4/\Intr_2$ orbifold, the fourth $\U{1}$ arises by symmetry 
breaking of the hidden $\SU{8}'$ in the blow--up procedure.

Turning on additional fluxes \eqref{eqn:MSSMResolutionData2} on $E'_{r'}$ 
induces a further gauge symmetry breaking to 
\equ{
\label{eq:BlowupWithFluxGaugeGroup}
\SU{5} \times \SU{5}' \times \U{1}^8~. 
}
Since, these fluxes are located at resolved fixed points of the other orbifold 
twist $v_\go$, they respect a different six--dimensional supersymmetry. This 
means that by switching on these fluxes the model becomes $\mathcal{N}=1$ in 
four dimensions. However, because the divisors $E'_{r'}$ do not intersect with 
$E_r$ (i.e.\ the fixed tori of the $g_\gth$-- and $g_\go$--twisted sectors do 
not intersect) see figure \ref{fig:fixedtori0-2}, this does not enforce any 
chiral projection on the matter spectrum (in contrast to, say, the gravitino): 
The matter states on $K3$ (i.e.\ the blow--up of 
the intermediate $T^4/\Intr_2$) are simply decomposed into four--dimensional 
superfields and their representations are branched according to the symmetry 
breaking \eqref{eq:BlowupWithFluxGaugeGroup}.

\newpage 
{\small
\begin{center}
\begin{longtable}{|c|c|c|c|}
\caption{The first column gives 6D $\mathcal{N}=1$ multiplets on the $T^4/\Z{2}$ 
orbifold with twist $g_\gth$ and gauge embedding $V_\gth$ and $W_3$ from 
equation \eqref{eq:ShiftsandWilson}. The second 
column indicates which state is the blow--up mode and gives the field 
redefinitions necessary to match the orbifold and blow--up states. In the third 
column we only indicate the states which are part of the four--dimensional chiral spectrum, 
i.e.\ those for which $\tilde N_{4D}$, given in the last column, is positive. 
\label{tab:OrbifoldBlowUp}}
\\ 
\hline
6D $\mathcal{N}=1$ super              & Blow--up induced redefinitions of     & Surviving 4D                         & \!4D multi-\! \\
multiplet on $T^4/\Z{2}$              & its chiral superfield component(s)    & chiral superfields                   & plicity \\[.5ex]  
($\E{6}\times \SU{8}'\times \U{1}^3$) & ($\E{6}\times \SU{7}'\times \U{1}^4$) & ($\SU{5}\times\SU{5}'\times\U{1}^8$) & $\tilde N_{4D}$\\ 
\hline
\hline
\endfirsthead
\hline
6D $\mathcal{N}=1$ super              & Blow--up induced redefinitions of     & Surviving 4D                         & \!4D multi-\! \\
multiplet on $T^4/\Z{2}$              & its chiral superfield component(s)    & chiral superfields                   & plicity \\[.5ex]  
($\E{6}\times \SU{8}'\times \U{1}^3$) & ($\E{6}\times \SU{7}'\times \U{1}^4$) & ($\SU{5}\times\SU{5}'\times\U{1}^8$) & $\tilde N_{4D}$\\ 
\hline
\hline
\endhead
\hline
\multicolumn{4}{|r|}{continued \ldots}\\
\hline
\endfoot
\hline \endlastfoot
\multicolumn{4}{|c|}{untwisted gauge sector} \\ 
\hline
\hline
$\left( \rep{78}, \rep{1}\right)_{\left( 0, 0, 0\right)}$\TabSp &  $\left( \rep{78}, \rep{1}\right)_{\left( 0, 0, 0\right)}$    & $\left(\crep{10},  \rep{1}\right)_{\left( 0, 0, 0, 0, 1, 0,-3, 0\right)}$ &  6 \\
(vector)                                                        &                                                               & $\left(  \rep{5},  \rep{1}\right)_{\left( 0, 0, 0, 0, 0, 0,-6, 0\right)}$ &  6 \\
                                                                &                                                               & $\left(  \rep{1},  \rep{1}\right)_{\left( 0, 0, 0, 0, 2, 0, 0, 0\right)}$ &  6 \\

\hline 
$\left(  \rep{1},\rep{63}\right)_{\left( 0, 0, 0\right)}$\TabSp & $\left(  \rep{1}, \rep{48}\right)_{\left( 0, 0, 0, 0\right)}$ & $\left(  \rep{1}, \crep{5}\right)_{\left( 0, 0, 0, 0, 0, 1, 0, 7\right)}$ &  6 \\
(vector)                                                        &                                                               & $\left(  \rep{1}, \crep{5}\right)_{\left( 0, 0, 0, 0, 0,-1, 0, 7\right)}$ &  6 \\
\cline{2-4}
\TabSp                                                          & $\left(  \rep{1}, \crep{7}\right)_{\left( 0, 0, 0, 4\right)}$ & --                                                                        & -- \\
\cline{2-4}
                                                                & $\left(  \rep{1},  \rep{7}\right)_{\left( 0, 0, 0,-4\right)}$ & $\left(  \rep{1},  \rep{1}\right)_{\left( 0, 0, 0,-4, 0, 1, 0, 5\right)}$ &  6 \\
                                                                &                                                               & $\left(  \rep{1},  \rep{1}\right)_{\left( 0, 0, 0,-4, 0,-1, 0, 5\right)}$ &  6 \\
\cline{2-4}
                                                                & $\left(  \rep{1},  \rep{1}\right)_{\left( 0, 0, 0, 0\right)}$ & --                                                                        & -- \\
\hline
\hline
\multicolumn{4}{|c|}{untwisted matter sectors: $U_a$, $a=2,3$} \\ 
\hline
\hline
$\left( \rep{27}, \rep{1}\right)_{\left(-1, 0,-1\right)}$       & $\left( \rep{27},  \rep{1}\right)_{\left(-1, 0,-1, 0\right)}$ & $\left(  \rep{1},  \rep{1}\right)_{\left(-1, 0,-1, 0,-1, 0, 5, 0\right)}$ &  6 \\
\cline{2-4}
(hyper)\TabSp                                                   & $\left(\crep{27},  \rep{1}\right)_{\left( 1, 0, 1, 0\right)}$ & $\left(  \rep{5},  \rep{1}\right)_{\left( 1, 0, 1, 0, 0, 0, 4, 0\right)}$ &  6 \\
                                                                &                                                               & $\left( \crep{5},  \rep{1}\right)_{\left( 1, 0, 1, 0,-1, 0, 1, 0\right)}$ &  6 \\
\hline
$\left(  \rep{1},\rep{70}\right)_{\left( 0, 0, 0\right)}$\TabSp & $\left(  \rep{1},\crep{35}\right)_{\left( 0, 0, 0,-2\right)}$ & $\left(  \rep{1}, \crep{5}\right)_{\left( 0, 0, 0,-2, 0, 0, 0,-8\right)}$ &  6 \\
\cline{2-4}
(half--hyper)\TabSp                                             & $\left(  \rep{1}, \rep{35}\right)_{\left( 0, 0, 0, 2\right)}$ & $\left(  \rep{1},\crep{10}\right)_{\left( 0, 0, 0, 2, 0, 0, 0,-6\right)}$ &  6 \\
\hline
$\left(  \rep{1}, \rep{1}\right)_{\left( 1, 0,-3\right)}$       & $\left(  \rep{1},  \rep{1}\right)_{\left( 1, 0,-3, 0\right)}$ & $\left(  \rep{1},  \rep{1}\right)_{\left( 1, 0,-3, 0, 0, 0, 0, 0\right)}$ &  6 \\
\cline{2-4}
(hyper)                                                         & $\left(  \rep{1},  \rep{1}\right)_{\left(-1, 0, 3, 0\right)}$ & --                                                                        & -- \\
\hline 
$\left(  \rep{1}, \rep{1}\right)_{\left( 0, 2, 0\right)}$       & $\left(  \rep{1},  \rep{1}\right)_{\left( 0, 2, 0, 0\right)}$ & $\left(  \rep{1},  \rep{1}\right)_{\left( 0, 2, 0, 0, 0, 0, 0, 0\right)}$ &  6 \\
\cline{2-4}
(hyper)                                                         & $\left(  \rep{1},  \rep{1}\right)_{\left( 0,-2, 0, 0\right)}$ & --                                                                        & -- \\
\hline
\hline
\multicolumn{4}{|c|}{twisted matter sector at the fixed tori: $r=(0,n_4,n_5,0)$, $n_4, n_5 = 0, 1$} \\ 
\hline
\hline
$\left(\rep{1},\rep{8}\right)_{\left(-\tfrac{1}{2},-\tfrac{1}{2},-\tfrac{3}{2}\right)}$        & $\left(\rep{1}, \rep{1}\right)_{\left( \tfrac{1}{2}, \tfrac{1}{2}, \tfrac{3}{2},-\tfrac{7}{2}\right)} = e^{+b_r}$                                                          & blow--up mode  &  axion \\
\cline{2-4}
(hyper)                                                                                        & $\left(\rep{1}, \rep{1}\right)_{\left(-\tfrac{1}{2},-\tfrac{1}{2},-\tfrac{3}{2}, \tfrac{7}{2}\right)} = e^{+b_r} \left(\rep{1}, \rep{1}\right)_{\left( 0, 0, 0, 0\right)}$ & -- & -- \\ 
\cline{2-4}
                                                                                               & $\left(\rep{1}, \rep{7}\right)_{\left(-\tfrac{1}{2},-\tfrac{1}{2},-\tfrac{3}{2},-\tfrac{1}{2}\right)} = e^{+b_r} \left(\rep{1}, \rep{7}\right)_{\left( 0, 0, 0,-4\right)}$ & -- & -- \\
\cline{2-4}
\TabSp                                                                                         & $\left(\rep{1},\crep{7}\right)_{\left( \tfrac{1}{2}, \tfrac{1}{2}, \tfrac{3}{2}, \tfrac{1}{2}\right)} = e^{-b_r} \left(\rep{1},\crep{7}\right)_{\left( 0, 0, 0, 4\right)}$ & $\left(\rep{1}, \rep{1}\right)_{\left( 0, 0, 0, 4, 0, 1, 0,-5\right)}$ &  6 \\
                                                                                               &                                                                                                                                                                            & $\left(\rep{1}, \rep{1}\right)_{\left( 0, 0, 0, 4, 0,-1, 0,-5\right)}$ &  6 \\
\hline 
$\left(\rep{1},\rep{8}\right)_{\left(\tfrac{1}{2}, -\tfrac{1}{2},\tfrac{3}{2}\right)}$         & $\left(\rep{1}, \rep{1}\right)_{\left( \tfrac{1}{2},-\tfrac{1}{2}, \tfrac{3}{2}, \tfrac{7}{2}\right)} = e^{+b_r} \left(\rep{1}, \rep{1}\right)_{\left( 1, 0, 3, 0\right)}$ & -- & -- \\ 
\cline{2-4}
(hyper)                                                                                        & $\left(\rep{1}, \rep{1}\right)_{\left(-\tfrac{1}{2}, \tfrac{1}{2},-\tfrac{3}{2},-\tfrac{7}{2}\right)} = e^{-b_r} \left(\rep{1}, \rep{1}\right)_{\left(-1, 0,-3, 0\right)}$ & -- & -- \\ 
\cline{2-4}
\TabSp                                                                                         & $\left(\rep{1},\crep{7}\right)_{\left(-\tfrac{1}{2}, \tfrac{1}{2},-\tfrac{3}{2}, \tfrac{1}{2}\right)} = e^{+b_r} \left(\rep{1},\crep{7}\right)_{\left( 0, 1, 0,-3\right)}$ & $\left(\rep{1}, \rep{1}\right)_{\left( 0, 1, 0,-3, 0, 1, 0,-5\right)}$ &  6 \\
                                                                                               &                                                                                                                                                                            & $\left(\rep{1}, \rep{1}\right)_{\left( 0, 1, 0,-3, 0,-1, 0,-5\right)}$ &  6 \\                                                                                                    
\cline{2-4}
                                                                                               & $\left(\rep{1}, \rep{7}\right)_{\left( \tfrac{1}{2},-\tfrac{1}{2}, \tfrac{3}{2},-\tfrac{1}{2}\right)} = e^{-b_r} \left(\rep{1}, \rep{7}\right)_{\left( 0,-1, 0, 3\right)}$ & -- & -- \\
\hline
\newpage 
\hline
\multicolumn{4}{|c|}{twisted matter sector at the fixed tori:  $r=(0,n_4,n_5,1)$, $n_4, n_5 = 0, 1$} \\ 
\hline
\hline
$\left(\rep{1}, \rep{8}\right)_{\left(-\tfrac{1}{2},-\tfrac{1}{2},-\tfrac{3}{2}\right)}$       & $\left(\rep{1}, \rep{1}\right)_{\left(-\tfrac{1}{2},-\tfrac{1}{2},-\tfrac{3}{2}, \tfrac{7}{2}\right)} = e^{+b_r}$                                                          & blow--up mode  &  axion \\
\cline{2-4}
(hyper)                                                                                        & $\left(\rep{1}, \rep{1}\right)_{\left( \tfrac{1}{2}, \tfrac{1}{2}, \tfrac{3}{2},-\tfrac{7}{2}\right)} = e^{+b_r} \left(\rep{1}, \rep{1}\right)_{\left( 0, 0, 0, 0\right)}$ & -- & -- \\ 
\cline{2-4}
\TabSp                                                                                         & $\left(\rep{1},\crep{7}\right)_{\left( \tfrac{1}{2}, \tfrac{1}{2}, \tfrac{3}{2}, \tfrac{1}{2}\right)} = e^{+b_r} \left(\rep{1},\crep{7}\right)_{\left( 0, 0, 0, 4\right)}$ & $\left(\rep{1}, \rep{1}\right)_{\left( 0, 0, 0, 4, 0, 1, 0,-5\right)}$ &  6 \\
                                                                                               &                                                                                                                                                                            & $\left(\rep{1}, \rep{1}\right)_{\left( 0, 0, 0, 4, 0,-1, 0,-5\right)}$ &  6 \\
\cline{2-4}
                                                                                               & $\left(\rep{1}, \rep{7}\right)_{\left(-\tfrac{1}{2},-\tfrac{1}{2},-\tfrac{3}{2},-\tfrac{1}{2}\right)} = e^{-b_r} \left(\rep{1}, \rep{7}\right)_{\left( 0, 0, 0,-4\right)}$ & -- & -- \\ \hline
\hline
$\left(\rep{1}, \rep{8}\right)_{\left( \tfrac{1}{2},-\tfrac{1}{2}, \tfrac{3}{2}\right)}$       & $\left(\rep{1}, \rep{1}\right)_{\left(-\tfrac{1}{2}, \tfrac{1}{2},-\tfrac{3}{2},-\tfrac{7}{2}\right)} = e^{+b_r} \left(\rep{1}, \rep{1}\right)_{\left(-1, 0,-3, 0\right)}$ & -- & -- \\ 
\cline{2-4}
(hyper)                                                                                        & $\left(\rep{1}, \rep{1}\right)_{\left( \tfrac{1}{2},-\tfrac{1}{2}, \tfrac{3}{2}, \tfrac{7}{2}\right)} = e^{-b_r} \left(\rep{1}, \rep{1}\right)_{\left( 1, 0, 3, 0\right)}$ & -- & -- \\  \hline
\cline{2-4}
                                                                                               & $\left(\rep{1}, \rep{7}\right)_{\left( \tfrac{1}{2},-\tfrac{1}{2}, \tfrac{3}{2},-\tfrac{1}{2}\right)} = e^{+b_r} \left(\rep{1}, \rep{7}\right)_{\left( 0,-1, 0, 3\right)}$ & -- & -- \\
\cline{2-4}
\TabSp                                                                                         & $\left(\rep{1},\crep{7}\right)_{\left(-\tfrac{1}{2}, \tfrac{1}{2},-\tfrac{3}{2}, \tfrac{1}{2}\right)} = e^{-b_r} \left(\rep{1},\crep{7}\right)_{\left( 0, 1, 0,-3\right)}$ & $\left(\rep{1}, \rep{1}\right)_{\left( 0, 1, 0,-3, 0, 1, 0,-5\right)}$ &  6 \\
                                                                                               &                                                                                                                                                                            & $\left(\rep{1}, \rep{1}\right)_{\left( 0, 1, 0,-3, 0,-1, 0,-5\right)}$ &  6 \\
\hline
\hline
\multicolumn{4}{|c|}{twisted matter sector at the fixed tori:  $r=(1,n_4,n_5,0)$, $n_4, n_5 = 0, 1$} \\ 
\hline
\hline
 $\left(\rep{1}, \rep{8}\right)_{\left(-1, \tfrac{1}{2}, 0\right)}$                            & $\left(\rep{1}, \rep{1}\right)_{\left( 1,-\tfrac{1}{2}, 0,-\tfrac{7}{2}\right)} = e^{+b_r}$                                                          & blow--up mode   & axion \\
\cline{2-4}
(hyper)                                                                                        & $\left(\rep{1}, \rep{1}\right)_{\left(-1, \tfrac{1}{2}, 0, \tfrac{7}{2}\right)} = e^{+b_r} \left(\rep{1}, \rep{1}\right)_{\left( 0, 0, 0, 0\right)}$ & -- & -- \\ 
\cline{2-4}
                                                                                               & $\left(\rep{1}, \rep{7}\right)_{\left(-1, \tfrac{1}{2}, 0,-\tfrac{1}{2}\right)} = e^{+b_r} \left(\rep{1}, \rep{7}\right)_{\left( 0, 0, 0,-4\right)}$ & -- & -- \\ 
\cline{2-4}
\TabSp                                                                                         & $\left(\rep{1},\crep{7}\right)_{\left( 1,-\tfrac{1}{2}, 0, \tfrac{1}{2}\right)} = e^{-b_r} \left(\rep{1},\crep{7}\right)_{\left( 0, 0, 0, 4\right)}$ & $\left(\rep{1}, \rep{1}\right)_{\left( 0, 0, 0, 4, 0, 1, 0,-5\right)}$ &  6 \\
                                                                                               &                                                                                                                                                      & $\left(\rep{1}, \rep{1}\right)_{\left( 0, 0, 0, 4, 0,-1, 0,-5\right)}$ &  6 \\
\hline 
$\left(\rep{1}, \rep{8}\right)_{\left( 1, \tfrac{1}{2}, 0\right)}$                             & $\left(\rep{1}, \rep{1}\right)_{\left( 1, \tfrac{1}{2}, 0, \tfrac{7}{2}\right)} = e^{+b_r} \left(\rep{1}, \rep{1}\right)_{\left( 2, 0, 0, 0\right)}$ & $\left(\rep{1}, \rep{1}\right)_{\left( 2, 0, 0, 0, 0, 0, 0, 0\right)}$ &  6 \\
\cline{2-4}
(hyper)                                                                                        & $\left(\rep{1}, \rep{1}\right)_{\left(-1,-\tfrac{1}{2}, 0,-\tfrac{7}{2}\right)} = e^{-b_r} \left(\rep{1}, \rep{1}\right)_{\left(-2, 0, 0, 0\right)}$ & -- & -- \\ 
\cline{2-4}
\TabSp                                                                                         & $\left(\rep{1},\crep{7}\right)_{\left(-1,-\tfrac{1}{2}, 0, \tfrac{1}{2}\right)} = e^{+b_r} \left(\rep{1},\crep{7}\right)_{\left( 0,-1, 0,-3\right)}$ & -- & -- \\ 
\cline{2-4}
                                                                                               & $\left(\rep{1}, \rep{7}\right)_{\left( 1, \tfrac{1}{2}, 0,-\tfrac{1}{2}\right)} = e^{-b_r} \left(\rep{1}, \rep{7}\right)_{\left( 0, 1, 0, 3\right)}$ & $\left(\rep{1}, \rep{5}\right)_{\left( 0, 1, 0, 3, 0, 0, 0,-2\right)}$ &  6 \\
\hline
\hline
\multicolumn{4}{|c|}{twisted matter sector at the fixed tori:  $r=(1,n_4,n_5,1)$, $n_4, n_5 = 0, 1$} \\ 
\hline
\hline
$\left(\rep{1}, \rep{8}\right)_{\left(-1, \tfrac{1}{2}, 0\right)}$                             & $\left(\rep{1}, \rep{1}\right)_{\left(-1, \tfrac{1}{2}, 0, \tfrac{7}{2}\right)} = e^{+b_r}$                                                          & blow--up mode   &  axion \\
\cline{2-4}
(hyper)                                                                                        & $\left(\rep{1}, \rep{1}\right)_{\left( 1,-\tfrac{1}{2}, 0,-\tfrac{7}{2}\right)} = e^{+b_r} \left(\rep{1}, \rep{1}\right)_{\left( 0, 0, 0, 0\right)}$ & -- & -- \\ 
\cline{2-4}
\TabSp                                                                                         & $\left(\rep{1},\crep{7}\right)_{\left( 1,-\tfrac{1}{2}, 0, \tfrac{1}{2}\right)} = e^{+b_r} \left(\rep{1},\crep{7}\right)_{\left( 0, 0, 0, 4\right)}$ & $\left(\rep{1}, \rep{1}\right)_{\left( 0, 0, 0, 4, 0, 1, 0,-5\right)}$ &  6 \\
                                                                                               &                                                                                                                                                      & $\left(\rep{1}, \rep{1}\right)_{\left( 0, 0, 0, 4, 0,-1, 0,-5\right)}$ &  6 \\
\cline{2-4}
                                                                                               & $\left(\rep{1}, \rep{7}\right)_{\left(-1, \tfrac{1}{2}, 0,-\tfrac{1}{2}\right)} = e^{-b_r} \left(\rep{1}, \rep{7}\right)_{\left( 0, 0, 0,-4\right)}$ & -- & -- \\ 
\hline 
$\left(\rep{1}, \rep{8}\right)_{\left( 1, \tfrac{1}{2}, 0\right)}$                             & $\left(\rep{1}, \rep{1}\right)_{\left(-1,-\tfrac{1}{2}, 0,-\tfrac{7}{2}\right)} = e^{+b_r} \left(\rep{1}, \rep{1}\right)_{\left(-2, 0, 0, 0\right)}$ & -- & -- \\ 
\cline{2-4}
(hyper)                                                                                        & $\left(\rep{1}, \rep{1}\right)_{\left( 1, \tfrac{1}{2}, 0, \tfrac{7}{2}\right)} = e^{-b_r} \left(\rep{1}, \rep{1}\right)_{\left( 2, 0, 0, 0\right)}$ & $\left(\rep{1}, \rep{1}\right)_{\left( 2, 0, 0, 0, 0, 0, 0, 0\right)}$ &  6 \\
\cline{2-4}
                                                                                               & $\left(\rep{1}, \rep{7}\right)_{\left( 1, \tfrac{1}{2}, 0,-\tfrac{1}{2}\right)} = e^{+b_r} \left(\rep{1}, \rep{7}\right)_{\left( 0, 1, 0, 3\right)}$ & $\left(\rep{1}, \rep{5}\right)_{\left( 0, 1, 0, 3, 0, 0, 0,-2\right)}$ &  6 \\
\cline{2-4}
\TabSp                                                                                         & $\left(\rep{1},\crep{7}\right)_{\left(-1,-\tfrac{1}{2}, 0, \tfrac{1}{2}\right)} = e^{-b_r} \left(\rep{1},\crep{7}\right)_{\left( 0,-1, 0,-3\right)}$ & -- & -- \\ 
\hline
\end{longtable}
\end{center}
}

\subsubsection*{Generating four dimensional chirality}

States from $E_r$ feel the flux on their ``fixed torus'', 
i.e.\ on $R_1$, so that the $B_1$ flux induces chirality in four dimensions. 
Whether a state is part of the charged chiral spectrum is decided by the 
operator 
\equ{
\tilde N_{4D} = 2 H_{B_1}~. 
}
In general \cite{Cremades:2004wa,Abe:2009uz}, if $\tilde N_{4D}$ is positive 
for a chiral superfield $\Phi_{res}$, then $\tilde N_{4D}$ copies of 
$\Phi_\text{res}$ appear in the four--dimensional spectrum. While, if 
$\tilde N_{4D}$ is negative $\Phi_\text{res}$ is completely projected out. 
Thus this relation (similarly to \eqref{Nfactorization}) shows that four dimensional 
chirality only arises if the flux $B_1$ is switched on.

Two important observations are in order: Chiral multiplets originating from the 
six--dimensional vector multiplets get an extra factor $(-1)$ in order to 
account for the different chiralities of vector and hypermultiplets in six 
dimensions. In addition, note that the effect of the Wilson line $W_1$ in the 
presence of a flux $B_1$ in the same torus can be seen in a field theoretical 
approach as a shift in the wave--functions. Hence, concerning the spectrum of 
massless modes it can be neglected.

As the 16 fixed points of $T^4/\Z{2}$ are identified pairwise by the $\Z{2}$ 
action of $g_\go$, one has to restrict to twisted states with $n_5 = 0$. 
Furthermore, $g_\go$ projects out all states from the untwisted sector $U_2$ as 
can be seen in the full orbifold model 
$T^6/\Intr_2\times \Intr_\text{2,rototrans}$. The result of the additional 
fluxes is listed in the third and fourth columns of Table 
\ref{tab:OrbifoldBlowUp}. The chiral part of the resulting spectrum agrees with 
the spectrum of the smooth model listed in Table \ref{tab:ExampleMagnetizedSpectrum}.

\section{Towards an CFT description of orbifolds with magnetized tori} 
\label{sc:OrbifoldCFT} 

In this section we propose modifications to the standard CFT construction of 
heterotic orbifolds in the presence of magnetized tori. To facilitate this 
discussion we first recall a few standard facts of heterotic orbifolds, i.e.\ 
orbifolds without any magnetic flux supported on the two--tori, $B_a=0$. 

\subsection{Standard modular invariance conditions}
\label{sc:ModInvStandard}

The conditions of modular invariance are compatible with the local Bianchi 
identities in the absence of $B_a$--fluxes in the following sense: If we choose 
space group elements $g=h=g_r$ or $g_{r'}$, as defined in equation 
\eqref{FPelements} we see from \eqref{ModInv} that the associated local shifts 
$V_{g_r}$ and $V_{g_{r'}}$ fulfill
\equ{
V_{g_r}^2 \equiv \frac 32~, 
\qquad 
V_{g_{r'}}^2 \equiv \frac 32~, 
\label{localModInv}
}
where $r, r'$ label the 8+8 fixed points of the twisted sectors of $\gth$ and 
$\go$, respectively. On the other hand, in the smooth picture if we assume 
gauge fluxes $V_r \cong V_{g_r}$ and $V'_{r'} \cong V_{g_{r'}}$ of 
length--square $3/2$ at all 8+8 resolved fixed points $r$ and $r'$, 
respectively, then modular invariance corresponds (modulo integers) to $1/8$th 
of the Bianchi identities \eqref{K3likeBIs} with $B_a=0$. 

\subsection{Heterotic description of the Schoen orbifold with magnetized tori}
\label{sc:ModInvTorusFlux}

Inspired by the logic put forward in \cite{Aldazabal:1997wi} we propose how the 
modular invariance conditions \eqref{ModInv} are modified in the presence of 
magnetically charged tori, $B_a \neq 0$. Since the magnetic fluxes are constant 
over the tori, it is natural to assume that at a given fixed point they only 
contribute as one over the number of fixed points, i.e. $1/8$. As can be 
inferred from the local Bianchi identities \eqref{K3likeBIs} the magnetic 
fluxes, $B_a$, contribute to the energy ($12$ is replaced by 
$12 + 2B_a \cdot B_3$, $a=1,2$). Hence, we propose that the local modular invariance 
conditions \eqref{localModInv} are modified to 
\equ{
V_{g_r}^2 \equiv \frac 32 + \frac 1{4}\, B_2 \cdot B_3~, 
\qquad 
V_{g_{r'}}^2 \equiv \frac 32 + \frac 1{4}\, B_1\cdot B_3~, 
\label{localModInvFlux}
}
In order to satisfy the quantization conditions \eqref{Quant} and the DUY 
equations in blow--down \eqref{BlowDownDUY} it is convenient to expand $B_3$ 
as a linear combination of $B_1$ and $B_2$ with negative coefficients. 
According to equation \eqref{localModInvFlux} this reduces the lengths of the 
local shifts $V_{g_r}$ and $V_{g_{r'}}$. For example, using e.g. 
$B_3 = -B_1 - B_2$ yields $V_{g_r}^2 \equiv \frac 32 - \frac 1{4}\, B_2^2$.

However, as we have seen in blow--up not only the consistency conditions, i.e.\ 
the Bianchi conditions, get modified in the presence of $B_a$--fluxes, but also 
the spectra. Therefore, one could imagine that the mass shell condition 
\eqref{MLR2} on orbifolds is modified as well when $B_a \neq 0$. In analogy to 
the proposal in \cite{Aldazabal:1997wi}, we expect that the left--moving mass 
is modified to 
\equ{
M_L^2 = \frac 12\, (P+V_{g_r})^2 + \tN -  \frac 34 - \frac 1{8}\, B_2 \cdot B_3~, 
\qquad 
M_L^2 = \frac 12\, (P+V_{g_{r'}})^2 + \tN - \frac 34 - \frac 1{8}\, B_1\cdot B_3~. 
\label{MLR2Flux} 
}
If we follow the interpretation of the local line bundle vectors as the shifted 
momenta \eqref{Psh} of twisted states that 
generate the blow--up at $r$ or $r'$ these equations will contribute new 
twisted states as blow--up modes, which where not part of the $B_a=0$ orbifold 
spectrum.

When one considers the standard heterotic orbifold, (massless) states, that survive the 
level matching condition, are subject to the orbifold projection conditions 
\eqref{Projections}. Modifications of these projections are, as far as we are 
aware, not discussed in the literature. Moreover, since it is unknown how the 
heterotic string is quantized in the presence of magnetized tori, there is also 
not an obvious computation that would determine the appropriate corrections. 
However, as usual we expect that at least self--projections, i.e.\ taking 
$h=g$, should not project out any state. Hence, at least the self--projection 
condition should be modified to
\equ{
V_g \cdot P_\text{sh} - v_g \cdot \left(p_\text{sh} + \Delta\tilde{N}_g\right) \equiv \frac 12 \big( V_g^2 - v_g^2  + \sfrac 14\, B_a\cdot B_3 \big)~, 
\label{SeflProjectionsFlux}
}
where $a=1$ for $g = g_{r'}$ and $a=2$ for $g=g_r$.

\subsection{Sample model as blow--up of orbifold with magnetized tori}

To illustrate our proposal we return to our example of an eight generation SU(5) GUT model discussed in Subsection \ref{sc:aGUT}. Notice, that the bundle vectors $V_r$ defined in \eqref{SampleInput} can be interpreted as the shifted left--moving momenta $P_\text{sh}$ of twisted states without oscillator excitations 
of a conventional $\Intr_2$ orbifold, since $V_r^2 = 3/2$ is interpreted as the 
masslessness condition $P_\text{sh}^2 = 3/2$. The bundle vectors $V'_{r'}$ on the 
other hand have ${V'_{r'}}^2 = 1/2$. In a conventional orbifold model these 
would correspond to twisted states with oscillators. However, as discussed in 
Section \ref{sc:OrbifoldCFT}, we expect that the left--moving mass formula gets 
modified to \eqref{MLR2Flux} in the presence of magnetized tori. If correct, 
one still interprets the $V'_{r'}$ as shifted left--moving momenta of twisted 
states without oscillators. Hence, even though this model has a blow--down 
limit, the resulting theory in this limit is not a conventional orbifold CFT.

\bibliographystyle{paper}
{\small

\begin{thebibliography}{10}

\bibitem{Faraggi:1991jr}
A.~E. Faraggi ``A new standard - like model in the four-dimensional free
  fermionic string formulation'' {\em Phys. Lett.} {\bf B278} (1992)
131--139.

\bibitem{Cleaver:1998sa}
G.~B. Cleaver, A.~E. Faraggi, and D.~V. Nanopoulos ``String derived {MSSM} and
  {M}-theory unification'' {\em Phys. Lett.} {\bf B455} (1999) 135--146
\href{http://www.arXiv.org/abs/hep-ph/9811427}{[{\tt arXiv:hep-ph/9811427}]}.

\bibitem{Dijkstra:2004cc}
T.~P.~T. Dijkstra, L.~R. Huiszoon, and A.~N. Schellekens ``{Supersymmetric
  Standard Model Spectra from RCFT orientifolds}'' {\em Nucl. Phys.} {\bf B710}
  (2005) 3--57
\href{http://www.arXiv.org/abs/hep-th/0411129}{[{\tt arXiv:hep-th/0411129}]}.

\bibitem{Dijkstra:2004ym}
T.~P.~T. Dijkstra, L.~R. Huiszoon, and A.~N. Schellekens ``{Chiral
  supersymmetric standard model spectra from orientifolds of Gepner models}''
  {\em Phys. Lett.} {\bf B609} (2005) 408--417
\href{http://www.arXiv.org/abs/hep-th/0403196}{[{\tt arXiv:hep-th/0403196}]}.

\bibitem{Dixon:1985jw}
L.~J. Dixon, J.~A. Harvey, C.~Vafa, and E.~Witten ``Strings on orbifolds'' {\em
  Nucl. Phys.} {\bf B261} (1985)
678--686.

\bibitem{Dixon:1986jc}
L.~J. Dixon, J.~A. Harvey, C.~Vafa, and E.~Witten ``Strings on orbifolds. 2''
  {\em Nucl. Phys.} {\bf B274} (1986)
285--314.

\bibitem{Nilles:2011aj}
H.~P. Nilles, S.~Ramos-S{\'a}nchez, P.~K. Vaudrevange, and A.~Wingerter ``{The
  Orbifolder: A Tool to study the Low Energy Effective Theory of Heterotic
  Orbifolds}'' {\em Comput.Phys.Commun.} {\bf 183} (2012) 1363--1380
\href{http://www.arXiv.org/abs/1110.5229}{[{\tt arXiv:1110.5229}]}.

\bibitem{Lebedev:2006kn}
O.~Lebedev {\em et al.} ``{A mini-landscape of exact MSSM spectra in heterotic
  orbifolds}'' {\em Phys. Lett.} {\bf B645} (2007) 88--94
\href{http://www.arXiv.org/abs/hep-th/0611095}{[{\tt arXiv:hep-th/0611095}]}.

\bibitem{Lebedev:2008un}
O.~Lebedev, H.~P. Nilles, S.~Ramos-S{\'a}nchez, M.~Ratz, and P.~K. Vaudrevange
  ``{Heterotic mini-landscape. ({II}). Completing the search for {MSSM} vacua
  in a {Z}(6) orbifold}'' {\em Phys.Lett.} {\bf B668} (2008) 331--335
\href{http://www.arXiv.org/abs/0807.4384}{[{\tt arXiv:0807.4384}]}.

\bibitem{Nibbelink:2009sp}
S.~Groot~Nibbelink, J.~Held, F.~Ruehle, M.~Trapletti, and P.~K.~S. Vaudrevange
  ``Heterotic {Z6-II} {MSSM} orbifolds in blowup'' {\em JHEP} {\bf 03} (2009)
  005
\href{http://www.arXiv.org/abs/0901.3059}{[{\tt arXiv:0901.3059}]}.

\bibitem{Buchmuller:2012mu}
W.~Buchm{\"u}ller, J.~Louis, J.~Schmidt, and R.~Valandro ``{Voisin-Borcea Manifolds
  and Heterotic Orbifold Models}'' {\em JHEP} {\bf 1210} (2012) 114
\href{http://www.arXiv.org/abs/1208.0704}{[{\tt arXiv:1208.0704}]}.

\bibitem{Blaszczyk:2009in}
M.~Blaszczyk {\em et al.} ``{A Z2xZ2 standard model}'' {\em Phys. Lett.} {\bf
  B683} (2010) 340--348
\href{http://www.arXiv.org/abs/0911.4905}{[{\tt arXiv:0911.4905}]}.

\bibitem{Blaszczyk:2010db}
M.~Blaszczyk, S.~Groot~Nibbelink, F.~Ruehle, M.~Trapletti, and P.~K.~S.
  Vaudrevange ``{Heterotic {MSSM} on a resolved orbifold}'' {\em JHEP} {\bf 09}
  (2010) 065
\href{http://www.arXiv.org/abs/1007.0203}{[{\tt arXiv:1007.0203}]}.

\bibitem{Bouchard:2005ag}
V.~Bouchard and R.~Donagi ``{An SU(5) heterotic standard model}'' {\em
  Phys.Lett.} {\bf B633} (2006) 783--791
\href{http://www.arXiv.org/abs/hep-th/0512149}{[{\tt arXiv:hep-th/0512149}]}.

\bibitem{Schoen:1988}
C.~Schoen ``{On fiber products of rational elliptic surfaces with section}''
  {\em Math. Z.} {\bf 197} (1988) 177--199.

\bibitem{Donagi:2000zf}
R.~Donagi, B.~A. Ovrut, T.~Pantev, and D.~Waldram ``{Standard model bundles on
  nonsimply connected Calabi-Yau threefolds}'' {\em JHEP} {\bf 0108} (2001) 053
\href{http://www.arXiv.org/abs/hep-th/0008008}{[{\tt arXiv:hep-th/0008008}]}.

\bibitem{Donagi:2000fw}
R.~Donagi, B.~A. Ovrut, T.~Pantev, and D.~Waldram ``{Spectral involutions on
  rational elliptic surfaces}'' {\em Adv.Theor.Math.Phys.} {\bf 5} (2002)
  499--561
\href{http://www.arXiv.org/abs/math/0008011}{[{\tt arXiv:math/0008011}]}.

\bibitem{Donagi:2004ub}
R.~Donagi, Y.-H. He, B.~A. Ovrut, and R.~Reinbacher ``{The Spectra of heterotic
  standard model vacua}'' {\em JHEP} {\bf 0506} (2005) 070
\href{http://www.arXiv.org/abs/hep-th/0411156}{[{\tt arXiv:hep-th/0411156}]}.

\bibitem{Gomez:2005ii}
T.~L. Gomez, S.~Lukic, and I.~Sols ``{Constraining the Kahler moduli in the
  heterotic standard model}'' {\em Commun.Math.Phys.} {\bf 276} (2007) 1--21
\href{http://www.arXiv.org/abs/hep-th/0512205}{[{\tt arXiv:hep-th/0512205}]}.

\bibitem{Braun:2005bw}
V.~Braun, Y.-H. He, B.~A. Ovrut, and T.~Pantev ``A standard model from the
  {E(8) x E(8)} heterotic superstring'' {\em JHEP} {\bf 06} (2005) 039
\href{http://www.arXiv.org/abs/hep-th/0502155}{[{\tt arXiv:hep-th/0502155}]}.

\bibitem{Anderson:2011ns}
L.~B. Anderson, J.~Gray, A.~Lukas, and E.~Palti ``{Two Hundred Heterotic
  Standard Models on Smooth Calabi-Yau Threefolds}'' {\em Phys.Rev.} {\bf D84}
  (2011) 106005
\href{http://www.arXiv.org/abs/1106.4804}{[{\tt arXiv:1106.4804}]}.

\bibitem{Anderson:2012yf}
L.~B. Anderson, J.~Gray, A.~Lukas, and E.~Palti ``{Heterotic Line Bundle
  Standard Models}'' {\em JHEP} {\bf 1206} (2012) 113
\href{http://www.arXiv.org/abs/1202.1757}{[{\tt arXiv:1202.1757}]}.

\bibitem{Donaldson:1985}
S.~Donalson ``Anti-self-dual {Y}ang--{M}ills connections over complex algebraic
  surfaces and stable vector bundles'' {\em Proc. Londan Math. Soc.} {\bf 50}
  (1985) 1--26.

\bibitem{Uhlenbeck:1986}
K.~Uhlenbeck and S.~Yau ``On the existence of {H}ermitian-{Y}ang-{M}ills
  connections in stable vector bundles'' {\em Comm. Pure and Appl. Math.} {\bf
  19} (1986) 257--293.

\bibitem{Blumenhagen:2005pm}
R.~Blumenhagen, G.~Honecker, and T.~Weigand ``Supersymmetric (non-)abelian
  bundles in the type {I} and {SO(32)} heterotic string'' {\em JHEP} {\bf 08}
  (2005) 009
\href{http://www.arXiv.org/abs/hep-th/0507041}{[{\tt arXiv:hep-th/0507041}]}.

\bibitem{Blumenhagen:2005ga}
R.~Blumenhagen, G.~Honecker, and T.~Weigand ``Loop-corrected compactifications
  of the heterotic string with line bundles'' {\em JHEP} {\bf 06} (2005) 020
\href{http://www.arXiv.org/abs/hep-th/0504232}{[{\tt arXiv:hep-th/0504232}]}.

\bibitem{Honecker:2006qz}
  G.~Honecker and M.~Trapletti,
  ``Merging Heterotic Orbifolds and K3 Compactifications with Line Bundles,''
  JHEP {\bf 0701} (2007) 051
  [hep-th/0612030].
  
\bibitem{Honecker:2007uw}
  G.~Honecker,
  ``Orbifolds versus smooth heterotic compactifications,''
  In *Karlsruhe 2007, SUSY 2007* 550-553
  [arXiv:0709.2037 [hep-th]].
  

\bibitem{Donagi:2008xy}
R.~Donagi and K.~Wendland ``{On orbifolds and free fermion constructions}''
  {\em J.Geom.Phys.} {\bf 59} (2009) 942--968
  \href{http://www.arXiv.org/abs/0809.0330}{[{\tt arXiv:0809.0330}]}.

\bibitem{Hebecker:2004ce}
  A.~Hebecker and M.~Trapletti,
  ``Gauge unification in highly anisotropic string compactifications,''
  Nucl.\ Phys.\ B {\bf 713} (2005) 173
  [hep-th/0411131].

\bibitem{Blumenhagen:2006ab}
  R.~Blumenhagen and E.~Plauschinn,
  ``Intersecting D-branes on shift Z(2) x Z(2) Orientifolds,''
  JHEP {\bf 0608} (2006) 031
  [hep-th/0604033].

\bibitem{Fischer:2012qj}
M.~Fischer, M.~Ratz, J.~Torrado, and P.~K. Vaudrevange ``{Classification of
  symmetric toroidal orbifolds}''
\href{http://www.arXiv.org/abs/1209.3906}{[{\tt arXiv:1209.3906}]}.

\bibitem{Konopka:2012gy}
S.~J. Konopka ``{Non Abelian orbifold compactifications of the heterotic
  string}''
\href{http://www.arXiv.org/abs/1210.5040}{[{\tt arXiv:1210.5040}]}.

\bibitem{Cremades:2004wa}
D.~Cremades, L.~Ib{\'a}{\~n}ez, and F.~Marchesano ``{Computing Yukawa couplings from
  magnetized extra dimensions}'' {\em JHEP} {\bf 0405} (2004) 079
  \href{http://www.arXiv.org/abs/hep-th/0404229}{[{\tt arXiv:hep-th/0404229}]}.

\bibitem{Abe:2009vi}
H.~Abe, K.-S. Choi, T.~Kobayashi, and H.~Ohki ``{Non-Abelian Discrete Flavor
  Symmetries from Magnetized/Intersecting Brane Models}'' {\em Nucl.Phys.} {\bf
  B820} (2009) 317--333 \href{http://www.arXiv.org/abs/0904.2631}{[{\tt
  arXiv:0904.2631}]}.

\bibitem{Abe:2009uz}
H.~Abe, K.-S. Choi, T.~Kobayashi, and H.~Ohki ``{Magnetic flux, Wilson line and
  orbifold}'' {\em Phys.Rev.} {\bf D80} (2009) 126006
  \href{http://www.arXiv.org/abs/0907.5274}{[{\tt arXiv:0907.5274}]}.

\bibitem{Bouchard:2007mf}
V.~Bouchard and R.~Donagi ``{On a class of non-simply connected Calabi-Yau
  threefolds}'' {\em Commun.Num.Theor.Phys.} {\bf 2} (2008) 1--61
\href{http://www.arXiv.org/abs/0704.3096}{[{\tt arXiv:0704.3096}]}.

\bibitem{Denef:2004dm}
F.~Denef, M.~R. Douglas, and B.~Florea ``{Building a better racetrack}'' {\em
  JHEP} {\bf 06} (2004) 034
\href{http://www.arXiv.org/abs/hep-th/0404257}{[{\tt arXiv:hep-th/0404257}]}.

\bibitem{Lust:2006zh}
D.~L{\"u}st, S.~Reffert, E.~Scheidegger, and S.~Stieberger ``{Resolved toroidal
  orbifolds and their orientifolds}'' {\em Adv. Theor. Math. Phys.} {\bf 12}
  (2008) 67--183
\href{http://www.arXiv.org/abs/hep-th/0609014}{[{\tt arXiv:hep-th/0609014}]}.

\bibitem{Nibbelink:2007rd}
S.~{Groot Nibbelink}, M.~Trapletti, and M.~Walter ``Resolutions of {$C^n/Z_n$}
  orbifolds, their {U(1)} bundles, and applications to string model building''
  {\em JHEP} {\bf 03} (2007) 035
\href{http://www.arXiv.org/abs/hep-th/0701227}{[{\tt arXiv:hep-th/0701227}]}.

\bibitem{Nibbelink:2007pn}
S.~{Groot Nibbelink}, T.-W. Ha, and M.~Trapletti ``{Toric Resolutions of
  heterotic orbifolds}'' {\em Phys. Rev.} {\bf D77} (2008) 026002
\href{http://www.arXiv.org/abs/0707.1597}{[{\tt arXiv:0707.1597}]}.

\bibitem{Nibbelink:2008tv}
S.~Groot~Nibbelink, D.~Klevers, F.~Pl{\"o}ger, M.~Trapletti, and P.~K.~S.
  Vaudrevange ``{Compact heterotic orbifolds in blow-up}'' {\em JHEP} {\bf 04}
  (2008) 060
\href{http://www.arXiv.org/abs/0802.2809}{[{\tt arXiv:0802.2809}]}.

\bibitem{Nibbelink:2010wm}
S.~Groot~Nibbelink ``{Heterotic orbifold resolutions as (2,0) gauged linear
  sigma models}'' {\em Fortsch.Phys.} {\bf 59} (2011) 454--493
\href{http://www.arXiv.org/abs/1012.3350}{[{\tt arXiv:1012.3350}]}.

\bibitem{Aldazabal:1997wi}
G.~Aldazabal, A.~Font, L.~E. Ib{\'a}{\~n}ez, A.~M. Uranga, and G.~Violero
  ``Non-perturbative heterotic {D = 6,4, N = 1} orbifold vacua'' {\em Nucl.
  Phys.} {\bf B519} (1998) 239--281
\href{http://www.arXiv.org/abs/hep-th/9706158}{[{\tt arXiv:hep-th/9706158}]}.

\end{thebibliography}

\providecommand{\href}[2]{#2}\begingroup\raggedright\endgroup

}

\end{document}